\documentclass[aps,pra,showpacs,groupaddress]{revtex4}
\usepackage{amssymb}
\usepackage{graphicx}
\usepackage{dcolumn}
\usepackage{bm}
\usepackage{amsmath}

\begin{document}

\title{Theory of Two-Photon Interactions with Broadband Down-Converted Light and Entangled Photons}
\author{Barak Dayan}
\affiliation{Norman Bridge Laboratory of Physics 12-33, California
Institute of Technology, Pasadena, California 91125, USA}

\begin{abstract}

When two-photon interactions are induced by down-converted light
with a bandwidth that exceeds the pump bandwidth, they can obtain
a behavior that is pulse-like temporally, yet spectrally narrow.
At low photon fluxes this behavior reflects the time and energy
entanglement between the down-converted photons. However,
two-photon interactions such as two-photon absorption (TPA) and
sum-frequency generation (SFG) can exhibit such a behavior even at
high power levels, as long as the final state (i.e. the atomic
level in TPA, or the generated light in SFG) is narrowband enough.
This behavior does not depend on the squeezing properties of the
light, is insensitive to linear losses, and has potential
applications. In this paper we describe analytically this behavior
for travelling-wave down-conversion with continuous or pulsed
pumping, both for high- and low-power regimes. For this we derive
a quantum-mechanical expression for the down-converted amplitude
generated by an arbitrary pump, and formulate operators that
represent various two-photon interactions induced by broadband
light. This model is in excellent agreement with experimental
results of TPA and SFG with high power down-converted light and
with entangled photons [Dayan $\textit{et al.}$, Phys. Rev. Lett.
$\textbf{93,}$ 023005 (2004), 
$\textbf{94}$, 043602, (2005), Pe'er $\textit{et
al.}$, Phys. Rev. Lett. $\textbf{94},$ 073601 (2005)].
\end{abstract}

\pacs{42.50.Ct, 42.65.Lm, 42.50.Dv, 32.80.Qk }

\maketitle



\section{Introduction} \label{Introduction}

Parametrically down-converted light
\cite{Giallorenzi@Tang_PR_1968,Byer@Harris_PR_1968,Hong@Mandel_PRA_1985,Mandel&Wolf_1995}
exhibits correlations between the down-converted modes, and has
been a primary source in experimental quantum-optics
\cite{{Mattle@Zeilinger_PRL_1996},{Bouwmeester@Zeilinger_Nature_1997},{Furusawa@Polzik_Science_1998},{Boschi@Popescu_PRL_1998},Jennewein@Zeilinger_PRL_2000,{Naik@Kwiat_PRL_2000},{Tittel@Gisin_PRL_2000}}.
At low photon-fluxes, these correlations are inherently
nonclassical, and are exhibited in the generation of entangled
photon-pairs
\cite{{Burnham@Weinberg_PRL_1970},Mandel_PRL_1982,{Kwiat@Chiao_PRA_1993},{Kwiat@Zeilinger_PRL_1995}}.
Two-photon interactions induced by entangled photons are expected
to exhibit nonclassical features, in particular a linear
dependence on the intensity
\cite{{Janszky@Yushin_PRA_1987},Banacloche_PRL_1989,{Javanainen@Gould_PRA_1990},{Fei@Teich_PRL_1997},{Saleh@Teich_PRL_1998},{Perina@Teich_PRA_1998},{Georgiades@Kimble_PRA_1999}},
as was observed with two-photon absorption (TPA)
\cite{Georgiades@Kimble_PRL_1995} and with sum-frequency
generation (SFG) \cite{Dayan@Silberberg_PRL_2005}. At higher
powers, the correlations between the down-converted modes are
better expressed as quadrature correlations. In this regime, the
nonclassical nature of the correlations lies in the fact that
their precision can exceed the vacuum shot-noise level,
a phenomenon that is named squeezing
\cite{{Mollow@Glauber_PR_I_1967},Mollow@Glauber_PR_II_1967,{Mandel_PRL_1982},{McNeil@Gardiner_PRA_1983},{Walls_Nature_1983},{Wu@Wu_PRL_1986},{Slusher@Potasek_PRL_1987},{Laurat@Fabre_PRA_2005},{Laurat@Coudreau_OPL_2006},{Takeno@Furusawa_OE_2007}},
and is expected to effect two-photon interactions even at
photon-fluxes which exceed the single photon regime
\cite{Janszky@Yushin_PRA_1987,{Ficek@Drummond_PRA_1991},{Gardiner@Parkins_PRA_1994},{Zhou@Swain_PRA_1996},{Turchette@Kimble_PRA_1998},{Georgiades@Kimble_PRA_1999}}.
However, while squeezing (i.e. the high precision of the
correlations) is easily destroyed (for example, by linear losses),
the remaining correlations still have a dramatic effect on
two-photon interactions. Specifically, when the down-converted
bandwidth is significantly larger than the pump bandwidth, the
down-converted light can induce two-photon interactions with the
same efficiency and sharp temporal behavior as ultrashort pulses,
while exhibiting high spectral resolution as that of the
narrowband pump
\cite{Abram@Dolique_PRL_1986,{Dayan@Silberberg_QPH_2003},{Dayan@Silberberg_PRL_2004},{Peer@Silberberg_PRL_2005}}.
Although at low photon fluxes these properties are a manifestation
of the time and energy entanglement between the down-converted
photons, at high powers exactly the same properties are exhibited
if the final state of the induced two-photon interaction is
narrowband enough \cite{Dayan@Silberberg_PRL_2004}. This effect
occurs since the narrow bandwidth of the final state
"post-selects" only the contribution of photon pairs that are
complementary in energy. Thus, in the high-power regime, these
seemingly nonclassical properties are completely described within
the classical framework (in fact, they are equivalent to the
correlations that exist between the transmitted signal and its key
in spread-spectrum communication systems
\cite{Peer@Friesem_JLT_2004}), and do not depend on the squeezing
degree of the down-converted light. This equivalence of broadband
down-converted light to coherent ultrashort pulses also implies an
ability to coherently control and shape the induced two-photon
interactions with pulse-shaping techniques, although the
down-converted light is neither coherent, nor pulsed. This effect
was demonstrated both in high power
\cite{Dayan@Silberberg_QPH_2003,{Dayan@Silberberg_PRL_2004}} and
with broadband entangled photons \cite{Peer@Silberberg_PRL_2005},
in which case it can be viewed as shaping of the second-order
correlation function of the entangled photons
$\texttt{g}^2(\tau)$. The fact that these unique properties can be
exhibited at high powers, with no dependence on nonclassical
features of the light (such as entanglement or squeezing), makes
this phenomena
both interesting and applicable \cite{Peer@Friesem_JLT_2004,{Peer@Friesem_PRA_2006}}.\\

In this paper we describe analytically this unique behavior of
two-photon interactions with broadband down-converted light.
First, in section \ref{motion}, we analytically solve the
equations of motion for broadband parametric down-conversion with
arbitrary pump (continuous or pulsed), obtaining a
non-perturbative solution that is valid both for low-power and for
high-power down-converted light. The derivation takes into account
the specific spectrum of the pump, assuming only that it is
significantly narrower than the down-converted spectrum. The
solution, namely the annihilation operators for the down-converted
fields, is represented as a function of the down-converted
spectrum, a property that is typically easy to calculate, estimate
or measure.\\

In section \ref{operators}, we formulate quantum-mechanical
operators that can represent multiphoton phenomena, specifically
TPA, SFG and coincidence events, induced by any broadband light.
By evaluating the expectation value of these operators, using the
annihilation operators derived in section \ref{motion}, we obtain
in section \ref{evaluation} a generic analytic expression for
two-photon interactions induced by broadband down-converted light,
as a function of the pump spectrum and the down-converted
spectrum. In section \ref{results} we derive specific expressions
for TPA, SFG and coincidence events induced by broadband
down-converted light, and analyze their temporal and spectral
properties. In section \ref{Summary} we give a brief summary of
our results and add some concluding remarks.\\

\section{Solving the equations of motion for broadband down-conversion with an arbitrary pump} \label{motion}

In our derivations we have chosen to use a continuous-variables
version of the formalism suggested by Huttner \textit{et al.}
\cite{Huttner@Ben-Aryeh_PRA_1990}. This formalism represents the
quantum fields in terms of space-dependent spectral mode operators
$a(\omega,z)$, instead of time-dependent momentum mode operators
$a(k,t)$. The advantage is that unlike momentum modes, temporal
modes are unchanged by a dielectric medium, reflecting the
physical fact that while the energy density of the fields depends
on the medium, the energy \textit{flux} does not. Accordingly, a
temporal periodicity, instead of a spatial periodicity is assumed
in the quantization process. We have performed the transition to
continuous variables (taking the temporal periodicity to infinity)
following the guidelines of the same procedure in momentum and
space \cite{Blow@Phoenix_PRA_1990}. Specifically, assuming a
single polarization and a single spatial mode we may write:
\begin{eqnarray} \label{eq8}
&& E(t,z)=E^+(t,z)+E^-(t,z)\ , \ \
\nonumber \\
&& E^+(t,z)= i \int_0^{\infty} d\omega \: \sqrt{\frac{\hbar
\omega}{4\pi c \varepsilon S}} \ \hat{a}(\omega,z)e^{-i\omega (t-z/c)} , \ \ 
E^-(t,z)=\left(E^+\left(t,z\right)\right)^*
\end{eqnarray}

\noindent where $S$ is the beam area, $\varepsilon$ is the
permittivity, and $\hat{a}^{\dag}(\omega,z), \ \hat{a}(\omega,z)$
are the slowly varying complex amplitudes of the creation and
annihilation operators of the electromagnetic field
\cite{Jeffers@Barnett_PRA_1993}. A major advantage of this
formalism is the relative convenience at which we can define a
momentum operator $\hat{G}(z)$ for the electromagnetic field in a
dispersive medium, and use it as the generator for space
propagation. The equations (and hence the solutions) derived in
the following resemble those attained by using the classical
Maxwell equations, or by using the Hamiltonian for time
propagation and replacing the spatial coordinate $z$ by $t c$
(see, for example, \cite{Yariv_1989}); however the derivation
presented here is a multimode, continuum-frequency one, enabling
us to take into account the specific spectrum of the pump and the
fact that each signal mode is coupled by the pump to multiple
idler modes, and
vice versa.\\

The momentum operator related to the second-order nonlinear
polarization is of the form
\cite{Shen_PR_1967,{Abram_PRA_1987},{Huttner@Ben-Aryeh_PRA_1990},{Caves@Crouch_JOSAB_1987}}:

\begin{eqnarray}
\label{G} \hat{G}^{nl}(z)= \int_0^\infty d\omega_p \int_0^\infty
d\omega \: \hbar \: \beta(\omega_p,\omega)
\left(\hat{a}_s(\omega,z)\hat{a}_i(\omega_p-\omega)\hat{a}_p^{\dag}(\omega_p,z)e^{-i\Delta
k(\omega_p,\omega) z}+H.C. \right) \:
, \nonumber \\
\beta(\omega_p,\omega)=\chi(\omega,\omega_p)\sqrt{\frac{\hbar\omega_p\omega(\omega_p-\omega)}{16\pi\varepsilon_0
c^3 S \textsl{n}_p (\omega_p) \textsl{n}_s(\omega) \textsl{n}_i
(\omega_p-\omega)}} \: ,
\end{eqnarray}

\noindent where the subscripts $s,i,p$  denote the signal, idler
and pump modes, respectively, $\textsl{n}_{s,i,p}(\omega_{s,i,p})$
denote the corresponding indices of refraction, and with:
\begin{eqnarray} \label{Dk}
\Delta k(\omega_p,\omega)=k_p(\omega_p)- k_s(\omega)-
k_i(\omega_p-\omega) \: .
\end{eqnarray}

The primary assumption throughout our calculations is that the
down-converted spectrum $\Delta_\textsc{dc}$ is considerably
broader than the pump bandwidth $\delta_p$:
\begin{eqnarray} \label{D>d}
\Delta_\textsc{dc}\gg\delta_p \ ,
\end{eqnarray}
\noindent where we take $\Delta_\textsc{dc}$ to denote the
bandwidth of the signal (or the idler) field (note that the signal
and the idler have the same bandwidth). Having assumed a
narrowband pump, we can safely neglect the dependence of $\chi$
(which is typically real) and $\Delta k$ on $\omega_p$, since
typically the crystal's nonlinear properties vary only at much
larger scales of frequencies, and write:
$\beta(\omega_p,\omega)=\beta(\omega)$, and $\Delta k
(\omega_p,\omega)=\Delta k (\omega)$. We thus write the equation
of motion for $\hat{a_s(\omega,z)}$ for undepleted, strong pump,
replacing $\hat{a}_p(\omega_p,z)$ with the spectral amplitude of
the classical pump field $\textsf{A}_p(\omega_p)$:

\begin{eqnarray}
\label{[aG]} \frac{ \partial \hat{a}_s( \omega , z ) }{\partial z
} &=& -\frac{i}{\hbar}\big[ \hat{a}_s ( \omega , z
),\hat{G}^{nl}(z)\big] \nonumber \\ &=& - i \beta(\omega) e^{-i
\Delta k (\omega) z} \int_0^\infty d\omega_p
\textsf{A}_p(\omega_p) \: \hat{a}_i^{\dag}(\omega_p-\omega,z)
\end{eqnarray}

\noindent The same can be formulated for the idler, which leads
to:

\begin{eqnarray}
\label{p_diff_OPA} \frac{\partial ^2 \:
\hat{a}_s(\omega,z)}{\partial z^2} &=&
|\beta(\omega)|^2\int^\infty_{-\infty} d\upsilon \int_0^\infty d
\omega_p \: \textsf{A}_p(\omega_p) \:
\textsf{A}_p^*(\omega_p+\upsilon) \: \hat{a}_s(\omega+\upsilon,z)
\nonumber \\ &-& \Delta k (\omega)\: \beta(\omega) e^{-i \Delta k
(\omega) z} \int_0^\infty d\omega_p \textsf{A}_p(\omega_p) \:
\hat{a}_i^{\dag}(\omega_p-\omega,z) \: ,
\end{eqnarray}

\noindent where $\upsilon=\widetilde\omega_P-\omega_P$. For a
stationary light field we can write
\cite{Jeffers@Loudon_PRA_1993}:
\begin{eqnarray}
\label{stat n} \int_0^\infty d \omega_p \: \textsf{A}_p(\omega_p)
\: \textsf{A}_p^*(\omega_p+\upsilon) = 2\pi I_p \:
\delta(\upsilon)
\end{eqnarray}

\noindent where $I_p = \langle
\texttt{I}_p(t)\rangle=\langle\big|A_p(t)\big|^2\rangle$ is the
mean power (in units of photon-flux) of the pump field, and so:
\begin{eqnarray}
\int^\infty_{-\infty} d\upsilon \int_0^\infty d \omega_p \:
\textsf{A}_p(\omega_p) \: \textsf{A}_p^*(\omega_p+\upsilon) \:
\hat{a}_s(\omega+\upsilon,z) = 2\pi I_p  \: \hat{a}_s(\omega,z) \
.
\end{eqnarray}

A similar result can be obtained, for non-stationary, pulsed light
with a final duration of $\tau_p$, by approximating:

\begin{eqnarray}
\label{pulse n} \int_0^\infty d \omega_p \: \textsf{A}_p(\omega_p)
\: \textsf{A}_p^*(\omega_p+\upsilon) &\approx& \int_{\tau_p}
\texttt{I}_p(t) \: dt \ \ when \ |\upsilon|<\frac{\pi}{\tau_p} \nonumber \\
&\approx& 0 \ \ \ \ \ \ \ \ \ \ otherwise \ .
\end{eqnarray}

This approximation reflects the fact that the spectral amplitude
of a finite signal can be considered roughly constant within
spectral slices that are smaller than $2\pi/\tau$, where $\tau$ is
the signal's duration. Accordingly, since (under the condition of
Eq. (\ref{D>d})) pulsed down-converted light has a duration which
is always equal or shorter than the pump pulse, its spectral
components may be regarded as constant over spectral slices that
are smaller than $2\pi/\tau_p$. For this reason it is safe to
approximate:

\begin{eqnarray}
\label{p_diff_pulse} \int^\infty_{-\infty} d\upsilon \int_0^\infty
d \omega_p \: \textsf{A}_p(\omega_p) \:
\textsf{A}_p^*(\omega_p+\upsilon) \: \hat{a}_s(\omega+\upsilon,z)
 &\approx& \int_{\tau_p} \texttt{I}_p(t) \: dt \:
\beta(\omega)^2\int^{\frac{\pi}{\tau_p}}_{-\frac{\pi}{\tau_p}}
d\upsilon \: \hat{a}_s(\omega+\upsilon,z) \nonumber \\ &\approx&
\int_{\tau_p} \texttt{I}_p(t) \: dt \:
\beta(\omega)^2\int^{\frac{\pi}{\tau_p}}_{-\frac{\pi}{\tau_p}}
d\upsilon \: \hat{a}_s(\omega,z) \nonumber \\ &=& 2\pi I_p
\beta(\omega)^2 \: \hat{a}_s(\omega,z)
\end{eqnarray}

\noindent obtaining the same result as with a stationary pump,
except for the fact that the averaging $I_p = \langle
\texttt{I}_p(t)\rangle$ is performed over the duration of the pump
pulse. Thus, both for stationary and non-stationary pumps, Eq.
(\ref{p_diff_OPA}) becomes:
\begin{eqnarray}
\label{p_diff_OPA approx} \frac{\partial ^2 \:
\hat{a}_s(\omega,z)}{\partial z^2} = 2\pi I_p \beta(\omega)^2 \:
\hat{a}_s(\omega,z) - \Delta k (\omega) \: \beta(\omega) e^{-i
\Delta k (\omega) z} \int_0^\infty d\omega_p
\textsf{A}_p(\omega_p) \: \hat{a}_i^{\dag}(\omega_p-\omega,z) \: .
\end{eqnarray}

The solution of Eqs. (\ref{[aG]}) and (\ref{p_diff_OPA approx})
is:

\begin{eqnarray}
\label{a_pumped} \hat{a}_s(\omega,z) \: e^{i\Delta k (\omega) z
/2} &=& \bigg( \cosh \big( \kappa(\omega) z \big) + \frac{i \Delta
k (\omega) }{2 \kappa(\omega)} \sinh  \big( \kappa(\omega) z \big)
\bigg) \: \hat{a}_s(\omega,0)  \nonumber \\ &-& \frac{i
\beta(\omega) }{\kappa(\omega)} \: \sinh  \big( \kappa(\omega) z
\big) \int_0^\infty d\omega_P \: \textsf{A}_p(\omega_p) \:
\hat{a}_i^{\dag}(\omega_p-\omega,0) \: .
\end{eqnarray}

\noindent with:

\begin{eqnarray} \label{KAPPA}
\kappa(\omega) =\sqrt{2\pi I_p \beta(\omega)^2 - \Delta
k(\omega)^2/4} \ .
\end{eqnarray}

The average photon flux spectral density for the signal and the
idler fields is therefore \cite{Blow@Phoenix_PRA_1990}:

\begin{eqnarray}
\label{n} \texttt{n}_{s,i}(\omega,z)&=&\frac{1}{2\pi}\int_0^\infty
d\omega' \big\langle
vac\big|\hat{a}_{s,i}^{\dag}(\omega,z)\hat{a}_{s,i}(\omega',z)\big|vac\big\rangle
\nonumber \\ &=& \frac{I_p \beta(\omega)^2}{\kappa(\omega^2)}
\sinh^2 \big( \kappa(\omega) z \big) \: .
\end{eqnarray}

Since the down-converted spectrum can readily be calculated or
measured for any specific down-conversion apparatus, we find it
convenient to present our following calculations using
$\texttt{n}_{s,i}(\omega)$ as \textit{given parameters}, thus
avoiding the issue of evaluating $\chi(\omega)$ and the other
elements of $\beta(\omega)$ that determine the down-converted
spectrum, and focusing on the behavior of two-photon interactions
induced by such a light. Assuming that the down-conversion process
occurred along a distance $L$, we denote:
\begin{eqnarray} \label{2pn}
n_{s,i}(\omega)= \frac{\kappa(\omega^2)}{I_p \beta(\omega)^2}
\texttt{n}_{s,i}(\omega,L) = \sinh^2 \big( \kappa(\omega) z \big)
\ ,
\end{eqnarray}
\noindent thus obtaining the following simple expression for
$\hat{a}_{s,i}(\omega,L)$:

\begin{eqnarray}
\label{a(L)} \hat{a}_{s,i}(\omega,L) \: e^{i\Delta k (\omega)L/2}
&=& \bigg( \sqrt{1+n_{s,i}(\omega)} + \frac{i \Delta k (\omega)
}{2 \kappa(\omega)}
\sqrt{n_{s,i}(\omega)}\:  \bigg) \: \hat{a}_{s,i}(\omega,0)  \nonumber \\
&-& \frac{i \beta(\omega) }{\kappa(\omega)} \:
\sqrt{n_{s,i}(\omega)} \int_0^\infty d\omega_P \:
\textsf{A}_p(\omega_p) \: \hat{a}_{i,s}^{\dag}(\omega_p-\omega,0)
\: .
\end{eqnarray}

For good phase-matching conditions (i.e $\Delta k(\omega) \ll
\sqrt{I_p}\beta(\omega)$), we get
$\kappa(\omega)\rightarrow\sqrt{2\pi I_p} \beta(\omega)$ and
therefore $\Delta k(\omega) / \kappa(\omega) \ll 1$, in which case
Eq. (\ref{a(L)}) can be further simplified to:

\begin{eqnarray}
\label{a(L)S} \hat{a}_{s,i}(\omega,L) \: e^{i\Delta k (\omega) z
/2} =\sqrt{1+n_{s,i}(\omega)} \: \hat{a}_{s,i}(\omega,0) -
\frac{i}{\sqrt{2\pi I_p}} \: \sqrt{n_{s,i}(\omega)} \int_0^\infty
d\omega_P \: \textsf{A}_p(\omega_p) \:
\hat{a}_{i,s}^{\dag}(\omega_p-\omega,0) \: .
\end{eqnarray}
 \noindent with:

\begin{eqnarray} \label{2pnS}
n_{s,i}(\omega)=2\pi\texttt{n}_{s,i}(\omega,L) \ ,
\end{eqnarray}

\section{Deriving operators for weak two-photon interactions induced by broadband fields}
\label{operators} In this section we derive expressions for weak
(perturbative) two-photon interactions induced by broadband
fields. Limiting ourselves to low efficiencies of interaction, we
neglect the depletion of the in-coming fields $a_{1,2}(\omega,z)$
that induce the interaction. Therefore we suppress in the
following their dependence on $z$, denoting
$a_{1,2}(\omega,z)=a_{1,2}(\omega,0)=a_{1,2}(\omega)$. We begin
with specific expressions for SFG and TPA, and then obtain a
generic expression which will be used in the following sections.

\subsection{SFG}

For SFG, we use the nonlinear momentum operator of Eq. (\ref{G}),
replacing $\hat{a}_p^\dag$ with the creation operator of the SFG
mode $\hat{a}_{\textsc{sfg}}^\dag$:
\begin{eqnarray}
\label{[SFGG]} \frac{ \partial \hat{a}_{\textsc{sfg}}( \Omega , z
) }{\partial z } &=& -\frac{i}{\hbar}\big[ \hat{a}_{\textsc{sfg}}
( \Omega, z ),\hat{G}^{nl}(z)\big] \nonumber \\ &=& -
i\int_0^\infty d\omega \: \beta(\omega,\Omega)
\hat{a}_1(\omega)\hat{a}_2(\Omega-\omega)e^{-i\Delta
k(\omega,\Omega) z} \: ,
\end{eqnarray}

\noindent with

\begin{eqnarray} \label{Dk sfg}
\Delta k(\Omega,\omega)=k_\textsc{sfg}(\Omega)- k_1(\omega)-
k_2(\Omega-\omega) \: .
\end{eqnarray}

This leads to the following approximation for
$\hat{a}_{\textsc{sfg}}( \Omega , z=L )$, where L is the overall
length of the nonlinear medium:
\begin{eqnarray}
\label{SFGG} \hat{a}_{\textsc{sfg}}( \Omega , L )&\approx&
\hat{a}_{\textsc{sfg}}( \Omega , 0) +\int_0^\infty d\omega \:
\left( \frac{e^{-i\Delta k(\omega,\Omega) L}-1} {\Delta
k(\omega,\Omega)} \right) \beta(\omega,\Omega)
\hat{a}_1(\omega)\hat{a}_2(\Omega-\omega) \nonumber \\
&=& \hat{a}_{\textsc{sfg}}( \Omega , 0) -i L \int_0^\infty d\omega
\: e^{-i\Delta k(\omega,\Omega) L/2} \: \textrm{sinc}\big[ \Delta
k(\omega,\Omega) L/2\big] \: \beta(\omega,\Omega) \:
\hat{a}_1(\omega)\hat{a}_2(\Omega-\omega) \: ,
\end{eqnarray}
\noindent where $\textrm{sinc}(x)=\sin(x)/x$. Since we are
interested only in the nonlinearly generated amplitude, we shall
ignore the first term, $\hat{a}_{\textsc{sfg}}( \Omega , 0)$, in
Eq. (\ref{SFGG}). The overall SFG photon flux through the plane
$z=L$ for a given initial light state $| \psi \rangle$ is
therefore:
\begin{eqnarray}
\label{nSFG} \texttt{N}_\textsc{sfg}(t,L)&=&\frac{1}{2\pi}
\int_0^\infty d\Omega \int_0^\infty d\Omega' \langle \psi |
\hat{a}_\textsc{sfg}^{\dag}(\Omega,L)e^{i\Omega t}
\hat{a}_\textsc{sfg} (\Omega',L)e^{-i\Omega' t} | \psi \rangle \nonumber \\
&=& \langle \psi | \hat{\eta}^\dag_\textsc{sfg}(t,L) \:
\hat{\eta}_\textsc{sfg}(t,L) | \psi \rangle
\end{eqnarray}

\noindent where $\hat{\eta}_\textsc{sfg}(t,L)$ is the photon flux
amplitude operator:
\begin{eqnarray} \label{eta}
\hat{\eta}_\textsc{sfg}(t,L)&=&\frac{1}{\sqrt{2\pi}}\int_0^\infty
d\Omega \: \hat{a}_\textsc{sfg}(\Omega,L)  \: e^{-i\Omega t} \nonumber \\
&=&-\frac{iL}{\sqrt{2\pi}}\int_0^\infty d\Omega \: e^{-i\Omega t}
\: \int_0^\infty d\omega \: e^{-i\Delta k(\omega,\Omega) L/2} \:
\textrm{sinc}\big[ \Delta k(\omega,\Omega) L/2\big] \:
\beta(\omega,\Omega) \:
\hat{a}_1(\omega)\hat{a}_2(\Omega-\omega) \nonumber \\
&=&-\frac{iL}{\sqrt{2\pi}}\int_0^\infty d\Omega \: e^{-i\Omega t}
\: \int_0^\infty d\omega \: \Phi(\omega,\Omega) \:
\hat{a}_1(\omega)\hat{a}_2(\Omega-\omega) \: ,
\end{eqnarray}
\noindent taking $\Phi(\omega,\Omega)$ to include the coupling
coefficient $\beta(\omega,\Omega)$ and phase matching terms:
\begin{eqnarray}
\label{phasematch} \Phi(\omega,\Omega)\equiv e^{-i\Delta
k(\omega,\Omega) L/2} \: \textrm{sinc}\big[ \Delta
k(\omega,\Omega) L/2\big] \: \beta(\omega,\Omega) \: .
\end{eqnarray}

Using Taylor expansion about the center frequency of the SFG
spectrum $\Omega_0$, $\Delta k (\omega,\Omega)$ can be separated
into two terms, where one depends on $\Omega-\Omega_0$ and the
other on $\omega-\langle \omega \rangle$:
\begin{eqnarray} \label{Dk 2 terms}
\Delta k (\omega,\Omega) &\approx&  \Big( \frac{\partial
k_\textsc{sfg} (\Omega)}{\partial \Omega} - \frac{\partial k_2
(\Omega-\omega)}{\partial \Omega} \Big) (\Omega-\Omega_0 ) + \Big(
\frac{\partial^2 k_\textsc{sfg} (\Omega)}{\partial \Omega^2} -
\frac{\partial^2 k_2 (\Omega-\omega)}{\partial \Omega^2} \Big)
(\Omega-\Omega_0 )^2  \nonumber \\ &+& \Big( \frac{\partial k_2
(\Omega-\omega)}{\partial \omega} - \frac{\partial k_1
(\omega)}{\partial \omega}  \Big) (\omega - \langle \omega
\rangle) +  \Big( \frac{\partial^2 k_2 (\Omega-\omega)}{\partial
\omega^2} - \frac{\partial^2 k_1
(\omega)}{\partial \omega^2}  \Big) (\omega - \langle \omega \rangle)^2 \nonumber \\
&=& \Delta k (\Omega, \langle \omega \rangle) + \Delta k (\omega,
\Omega_0) \ ,
\end{eqnarray}

\noindent where the first two terms represents the difference
between the group-velocities and between the group-velocity
dispersions of the pump and the idler, and the second represents
the difference between the group-velocities and between the
group-velocity dispersions of the signal and the idler. In type-I
phase-matching this implies a linear dependence on $\omega_p$
versus a much weaker quadratic dependence on $\omega$, since the
group-velocities of the signal and the idler are identical:

\begin{eqnarray} \label{Dk 2 terms type I}
\Delta k (\omega,\Omega) &\approx&  \Big( \frac{\partial
k_\textsc{sfg} (\Omega)}{\partial \Omega} - \frac{\partial k_2
(\Omega-\omega)}{\partial \Omega} \Big) (\Omega-\Omega_0 ) + \Big(
\frac{\partial^2 k_2 (\Omega-\omega)}{\partial \omega^2} -
\frac{\partial^2 k_1 (\omega)}{\partial \omega^2}  \Big) (\omega -
\langle \omega \rangle)^2  \ .
\end{eqnarray}

Since the $\textrm{sinc}$ function in Eqs.
(\ref{SFGG})-(\ref{phasematch}) results from integration over the
exponent $e^{i\Delta k}$, and since $\beta(\omega,\Omega)$ depends
very weakly on $\Omega$, for good phase-matching conditions (i.e.
small $\Delta k$), the approximation $\Delta k (\omega,\Omega)
\approx \Delta k (\Omega, \langle \omega \rangle) + \Delta k
(\omega, \Omega_0)$ enables us to represent the dependence of
$\Phi(\omega,\Omega)$ as:

\begin{eqnarray}
\label{g f sfg} && \Phi(\omega,\Omega) \approx
g_\textsc{sfg}(\Omega-\Omega_0)\:f_\textsc{sfg}(\omega,\Omega_0)
\: ,
\end{eqnarray}
with
\begin{eqnarray}
\label{g f sfg2} && g_\textsc{sfg}(\Omega-\Omega_0)=e^{-i\Delta
k(\Omega-\Omega_0) L/2} \: \textrm{sinc}\big[ \Delta
k(\Omega-\Omega_0) L/2\big] \: \nonumber
\\ && f_\textsc{sfg}(\omega,\Omega_0)=e^{-i\Delta
k(\omega,\Omega_0) L/2} \: \textrm{sinc}\big[ \Delta
k(\omega,\Omega_0) L/2\big] \: \beta(\omega,\Omega_0) \: ,
\end{eqnarray}

\noindent and with $\Omega_0$ being the center frequency of the
SFG spectrum. Assigning $\xi\equiv\Omega-\Omega_0$ we then rewrite
the photon flux amplitude operator (Eq. (\ref{eta})) as:
\begin{eqnarray} \label{eta2}
\hat{\eta}_\textsc{sfg}(t,L)&=&-\frac{iL\: e^{-i\Omega_0
t}}{\sqrt{2\pi}}\int d\xi \:  e^{-i\xi t} \: g_\textsc{sfg}(\xi)
\int_0^\infty d\omega \: f_\textsc{sfg}(\omega,\Omega_0) \:
\hat{a}_1(\omega)\hat{a}_2(\Omega_0+\xi-\omega) \: .
\end{eqnarray}

By applying Eq. (\ref{nSFG}), using $\hat{\eta}_\textsc{sfg}$ as
defined in Eq. (\ref{eta2}), we may evaluate the SFG intensity
induced by any initial state $| \psi \rangle$ of the light (note
again that this expression is valid only as long as the
up-conversion does not deplete the incoming fields).

\subsection{TPA}

Using second-order perturbation theory, very similar expressions
can be derived for TPA. The interaction Hamiltonian of an atom and
one spatial mode of the electromagnetic field takes the form of:
\begin{eqnarray} \label{HITPA} H_I^\textsc{tpa}(t) =
i \sum_{j}\sum_{k}\mu_{kj} \: |k\rangle \langle j | \:
e^{i\omega_{kj}t+\gamma_k t -\gamma_j t} \int_0^{\infty} d\omega
\: \sqrt{\frac{\hbar \omega}{4\pi c \varepsilon S}} \:  \Big[
\hat{a}(\omega) e^{-i\omega t} - \hat{a}^\dag(\omega) e^{i\omega
t)} \Big] \: ,
\end{eqnarray}
where $\mu_{kj}=\langle k | \mu | j \rangle$ are the dipole moment
matrix elements, $\omega_{kj}=\omega_k-\omega_j$,
$\gamma_k,\gamma_j$ are the level life times, and the summation is
performed over all the combinations of the unperturbed atomic
levels $|j\rangle, |k\rangle$. In order to evaluate the TPA
amplitude, we may use the second-order approximation for the
time-evolution operator that corresponds to this interaction
Hamiltonian, taking only the terms that contribute to a transition
from the initial (ground) level $|g\rangle$ to the final level
$|f\rangle$. Assuming the atom is initially in the ground state
$|g\rangle$, the probability $P_f^\textsc{tpa}$ for a light state
$| \psi \rangle$ to induce TPA can be represented as:
\begin{eqnarray} \label{Pf}
P_f^\textsc{tpa}(t)=\langle \psi | \: \hat{\eta}_\textsc{tpa}^\dag
(t) \: \hat{\eta}_\textsc{tpa}(t) \: | \psi \rangle \: ,
\end{eqnarray}

\noindent with $\hat{\eta}_\textsc{tpa}(t)$ defined in a very
similar way to Eq. (\ref{eta2}):

\begin{eqnarray} \label{etaTPA}
\hat{\eta}_\textsc{tpa}(t) &=& -\frac{e^{-i \omega_f t}} {4 \pi c
\varepsilon S \hbar} \sum_{n}\mu_{fn}\mu_{ng} \int_0^{\infty}
d\omega \int_0^{\infty} d\omega' \: \frac{\sqrt{\omega \omega'} \:
e^{i(\omega_{fg}-\omega-\omega')t} }
{(\omega_{fg}-\omega-\omega'-i\gamma_f)(\omega_{ng}-\omega-i\gamma_n)}
\: \hat{a}_1(\omega) \: \hat{a}_2(\omega')
 \nonumber \\
&=&- \frac{e^{-i \omega_f t}}{4 \pi c \varepsilon S \hbar}
\sum_{n}\mu_{fn}\mu_{ng} \int d\xi \frac{e^{-i\xi
t}}{\xi+i\gamma_f} \int_0^{\infty} d\omega \:  \frac{\sqrt{\omega
(\omega_{fg}+\xi-\omega)}} {(\omega_{ng}-\omega-i\gamma_n)} \:
\hat{a}_1(\omega) \: \hat{a}_2(\omega_{fg}+\xi-\omega)\: ,
\end{eqnarray}

\noindent where the subscripts $g,n,f$ denote the ground,
intermediate and final levels, respectively, and
$\xi=\omega+\omega'-\omega_{fg}$. Once again, the subscripts $1,2$
denote the two spatial modes of the fields that induce the
interaction. For convenience, we assume here that the atom is
located at $z_1=z_2=0$ along these modes. For non-resonant TPA,
i.e. when the spectra of the inducing fields do not overlap with
any resonant intermediate levels, the operator defined in Eq.
(\ref{etaTPA}) can approximated to:
\begin{eqnarray} \label{etanonresTPA}
\hat{\eta}_\textsc{tpa}(t)&\approx& - \frac{e^{-i \omega_{fg}
t}}{4 \pi c \varepsilon S \hbar} \frac{ \sum_{n}\mu_{fn}\mu_{ng} }
{(\omega_{ng}-\langle \omega \rangle )}\int d\xi \frac{e^{-i\xi
t}}{\xi+i\gamma_f} \int_0^{\infty} d\omega \: \sqrt{\omega
(\omega_{fg}+\xi-\omega)} \: \hat{a}_1(\omega) \:
\hat{a}_2(\omega_{fg}+\xi-\omega)\: ,
\end{eqnarray}
where $\langle \omega \rangle$ is the center frequency of the
field in mode $1$. \\

The similarity between Eq. (\ref{etanonresTPA}) and Eq.
(\ref{eta2}) becomes evident if we denote:
\begin{eqnarray} \label{TPA in gf}
&& \hat{\eta}_\textsc{tpa}(t)= \: constant \: \times e^{-i
\omega_{fg}t} \int d\xi \: e^{-i\xi t} \: g_\textsc{tpa}(\xi)
\int_0^{\infty} d\omega \: f_\textsc{tpa}(\omega,\omega_{fg}) \:
\hat{a}_1(\omega) \: \hat{a}_2(\omega_{fg}+\xi-\omega) \: ,
\end{eqnarray}
\noindent with:

\begin{eqnarray} \label{gfTPA}
&&g_\textsc{tpa}(\xi)=\frac{1}{\xi+i\gamma_f}  \nonumber \\
&&f_\textsc{tpa}(\omega,\omega_{fg}) = \sum_{n}\mu_{fn}\mu_{ng} \:
\frac{\sqrt{\omega (\omega_{fg}-\omega)}}
{(\omega_{ng}-\omega-i\gamma_n)} \: ,
\end{eqnarray}

\noindent and for non-resonant TPA :

\begin{eqnarray} \label{gf nrTPA}
&&g_\textsc{tpa}(\xi)=\frac{1}{\xi+i\gamma_f}  \nonumber \\
&&f_\textsc{tpa}(\omega,\omega_{fg}) =\sqrt{\omega
(\omega_{fg}-\omega)}\: ,
\end{eqnarray}

\noindent where in both expressions we approximated $\sqrt{\omega
(\omega_{fg}+\xi-\omega)}\approx\sqrt{\omega
(\omega_{fg}-\omega)}$, since the atomic level linewidth (which
defines the range over which $\xi$ is integrated) is negligible
compared to the optical frequencies.

\subsection{A generic two-photon operator}

In the next section we shall take advantage of the similarity
between the expression for SFG and non-resonant TPA, and perform
all the derivations using the following generic form for the
probability amplitude of the final state of the nonlinear
interaction:

\begin{eqnarray} \label{eta0}
\hat{\eta}(t,\tau_1,\tau_2)= e^{-i\Omega_0 t}\int d\xi \: e^{-i\xi
t} \: g(\xi) \int_0^{\infty} d\omega \: f(\omega,\Omega_0) \:
\hat{a}_1(\omega) e^{i\omega \tau_1}\:
\hat{a}_2(\Omega_0+\xi-\omega) e^{i(\Omega_0+\xi-\omega )\tau_2}
\: .
\end{eqnarray}

By introducing the exponents $e^{-i\omega_{1,2}\tau_{1,2}}$ we
take into account the possibility that the fields
$\hat{a}_{1,2}(\omega,0)$ have propagated freely separately,
accumulating temporal delays of $\tau_{1,2}$, respectively, before
inducing the interaction. The final probability for the
interaction, and hence the intensity $I(t,\tau_1,\tau_2)$ of the
measured signal is then represented by:
\begin{eqnarray} \label{I}
I(t,\tau_1,\tau_2) \propto  \langle \psi | \:
\hat{\eta}^\dag(t,\tau_1,\tau_2) \: \hat{\eta}(t,\tau_1,\tau_2) \:
| \psi \rangle \: .
\end{eqnarray}

The effect of any inhomogeneous broadening mechanism of the final
level may be taken into account by evaluating the intensity
$I(\Omega_0)$ of each homogeneously broadened subset with center
frequency $\Omega_0$, and defining:
\begin{eqnarray} \label{Itotal}
I^{total} \propto \int P(\Omega_0) I(\Omega_0) \: d\Omega_0 \: ,
\end{eqnarray}
\noindent with $P(\Omega_0)$ being the probability distribution of
the center frequency $\Omega_0$.\\

By assigning the appropriate expressions for
$g(\xi),f(\omega,\Omega_0)$,  Eqs. (\ref{eta0})-(\ref{Itotal}) may
represent SFG (using the definitions in Eq. (\ref{g f sfg})), TPA
(using Eq. (\ref{gfTPA})), or non-resonant TPA (using Eq. (\ref{gf
nrTPA})), as well as other two-photon interactions. For example,
to evaluate the rate of coincidences of photons at some optical
bandwidth $\Delta$ around $\Omega_0/2$, we may assign:
\begin{eqnarray} \label{gf coincidence}
f(\omega,\Omega_0) &\approx& \ 1 \ \ when \ |\omega-\Omega_0/2|<\Delta/2 \nonumber \\
&\approx& \ 0 \ \ \ \ \ \ \ \ \ \ otherwise, \nonumber \\
g(\xi) &\approx& \ 1 \ \ when \ |\xi|<\Delta \nonumber \\
&\approx& \ 0 \ \ \ \ \ \ \ \ \ \ otherwise,
\end{eqnarray}
\noindent where we assumed that the bandwidth $\Delta$ is smaller
than the optical frequency $\Omega_0/2$. Under these conditions,
$\hat{\eta}(t,\tau_1,\tau_2) \propto
\hat{E}_1^+(t-\tau_1)\hat{E}_2^+(t-\tau_2)$, and so the overall
intensity is simply proportional to the second-order correlation
function between the fields : $I(t,\tau_1,\tau_2) \propto
\texttt{g}^{(2)}(\tau_1-\tau_2)$, which is typically taken to
represent coincidence events.\\

Finally, $f(\omega,\Omega_0)$ may also represent any spectral
filters $\Theta(\omega)$ that are applied to the inducing light by
denoting:
\begin{eqnarray} \label{F=Theta}
f(\omega,\Omega_0)=\Theta(\omega)\Theta(\Omega_0-\omega) \ ,
\end{eqnarray}
where we assume that the spectral filtering is constant within
spectral slices that are narrower that the final state bandwidth,
and so we may neglect the dependence on $\xi$. If this is not the
case, than we should keep the dependence on $\xi$:
$f(\omega,\Omega_0,\xi)=\Theta(\omega)\Theta(\Omega_0+\xi-\omega)$
. In the case of SFG, a spectral filter can be applied to the
up-converted light as well, in which case its amplitude
transmission function should multiply $g(\xi)$.

\section{A generic expression for Two-photon interactions with broadband down-converted light: the coherent and incoherent
signals} \label{evaluation}


For the following derivations we shall assume that the signal,
idler and pump fields have each a single spatial mode and
polarization, that the down-conversion process occurred along a
distance $L$, and that the signal and idler may then travelled
along different paths, resulting in delays of $\tau_s,\tau_i$
respectively, before inducing the nonlinear interaction. Note that
in the following we consider only two-photon interactions that
result from cross-mixing of the signal and the idler fields and
not from self-mixing of the signal with itself or the idler with
itself (the possible contribution of such self-mixing terms will
be considered briefly later on). Assuming that good phase-matching
conditions are achieved for down-conversion at some bandwidth
$\Delta_\textsc{dc}$, we use the expressions obtained for
$\hat{a}_{s,i}(\omega,L)$ (Eq. (\ref{a(L)S})) in
$\hat{a}_{1,2}(\omega,0)$ of Eq. (\ref{eta0}), respectively:

\begin{eqnarray}
\label{e1} \hat{\eta}(t,\tau_s,\tau_i) &=& e^{-i\Omega_0 t}\int
d\xi \ g(\xi) \: e^{-i\xi t} \int_0^{\infty} d\omega \
f(\omega,\Omega_0) e^{i\omega \tau_s}
e^{i(\Omega_0+\xi-\omega) \tau_i} e^{-i\Delta k(\omega) L} \nonumber \\
&\times& \left( \sqrt{1+n_s(\omega)} \: \hat{a}_s(\omega) -
\frac{i}{\sqrt{2\pi I_p}} \: \sqrt{n_s(\omega)} \int_0^\infty
d\omega_P \: \textsf{A}_p(\omega_p) \:
\hat{a}_i^{\dag}(\omega_p-\omega)\right) \nonumber \\ &\times&
\Bigg( \sqrt{1+n_i(\Omega_0+\xi -\omega)} \:
\hat{a}_i(\Omega_0+\xi -\omega)  \nonumber \\ &-&
\frac{i}{\sqrt{2\pi I_p}} \: \sqrt{n_i(\Omega_0+\xi -\omega)}
\int_0^\infty d \omega'_P \: \textsf{A}_p(\omega'_p) \:
\hat{a}_s^{\dag}(\omega'_p+\omega-\Omega_0-\xi) \Bigg) \ .
\end{eqnarray}

\noindent Operating on the initial vacuum state and using
$a(\omega)a^\dag(\omega')=\delta(\omega-\omega')+a^\dag(\omega')a(\omega)$:

\begin{eqnarray}
\label{e2} \hat{\eta}(t,\tau_s,\tau_i)|vac\rangle &=&
e^{-i\Omega_0 t} \int d\xi \ g(\xi) \: e^{-i\xi t} \int_0^{\infty}
d\omega \: f(\omega,\Omega_0) e^{i\omega \tau_s}
e^{i(\Omega_0+\xi-\omega) \tau_i} e^{-i\Delta k(\omega) L} \nonumber \\
&\times& \Bigg( \frac{-1}{2\pi I_p} \:
\sqrt{n_s(\omega)n_i(\Omega_0+\xi -\omega)} \int_0^\infty
d\omega_P \: \textsf{A}_p(\omega_p)\int_0^\infty d \omega'_P \:
\textsf{A}_p(\omega'_p)  \nonumber \\ & \times &
\hat{a}_s^{\dag}(\omega'_p+\omega-\Omega_0-\xi)\hat{a}_i^{\dag}(\omega_p-\omega)
|vac\rangle \nonumber \\ & - & \frac{i}{\sqrt{2\pi
I_p}}\sqrt{\left(1+n_s(\omega) \right)n_i(\Omega_0+\xi -\omega)}
\: \textsf{A}_p(\Omega_0+\xi)\: |vac\rangle \Bigg) \ .
\end{eqnarray}

The intensity $I(t,\tau_s,\tau_i)$ can therefore be separated to
two components, which we will denote as the 'coherent' and the
'incoherent' signals:
\begin{eqnarray} \label{I=I+I} I(t,\tau_s,\tau_i) &\propto& \langle vac |
\hat{\eta}^\dag(t,\tau_s,\tau_i)\hat{\eta}(t,\tau_s,\tau_i)| vac
\rangle = I^c(t,\tau_s,\tau_i)+I^{ic}(t,\tau_s,\tau_i) \: ,
\end{eqnarray}

\noindent with:

\begin{eqnarray}
\label{IQ} I^c(t,\tau_s,\tau_i) &=& \frac{1}{2\pi I_p} \: \bigg|
\int d\xi \: g(\xi) \:  \textsf{A}_p(\Omega_0+\xi) \: e^{-i\xi
(t-\tau_i)} \int_0^{\infty} d\omega \: f(\omega,\Omega_0)
e^{-i\Delta k (\omega)L}
 \nonumber \\ &&\times \sqrt{\left(1+n_s(\omega)
\right)n_i(\Omega_0+\xi -\omega)} \: e^{-i\omega(\tau_i-\tau_s)} \
\bigg|^2 \: .
\end{eqnarray}

In order obtain $I^{ic}(t)$, we shall change variables in Eq.
(\ref{e2}): $u=\omega_p-\omega, \ v=\omega'_p +\omega-\xi$,
leading to:

\begin{eqnarray}
\label{c_alpha_vac} \hat{\eta}^{ic}(t,\tau_s,\tau_i)|vac\rangle
&=& \frac{e^{i\Omega_0 (\tau_i-t)}}{2 \pi I_p} \int_0^\infty du
\int_0^\infty dv \int d\xi \ g(\xi) \:e^{i\xi(\tau_i-t)}
\int_0^\infty d\omega_p
\: e^{i(\omega_p-u)(\tau_s-\tau_i)} \nonumber \\
&\times& f(\omega_p-u,\Omega_0) e^{-i \Delta k (\omega_p-u) L}
\sqrt{n_s(\omega_p-u)n_i(\Omega_0+\xi-\omega_p+u)} \nonumber \\
&\times& \textsf{A}_p(\omega_p) \: \textsf{A}_p(v+u+\xi-\omega_p)
\: \hat{a}_s^{\dag}(v-\Omega_0) \: \hat{a}_i^{\dag}(u) \:
|vac\rangle \: ,
\end{eqnarray}

\noindent and so:

\begin{eqnarray}
\label{IC} I^{ic}(t,\tau_s,\tau_i) &=& \frac{1}{(2 \pi I_p)^2}
\int_0^\infty du \int_0^\infty dv \: \bigg| \: \int d\xi \
g(\xi)\:e^{i\xi(\tau_i-t)} \int_0^\infty d\omega_p
\:\sqrt{n_s(\omega_p-u)n_i(\Omega_0+\xi-\omega_p+u)} \nonumber \\
&\times& f(\omega_p-u,\Omega_0)  e^{-i \Delta k (\omega_p-u) L}\:
\textsf{A}_p(\omega_p) \: \textsf{A}_p(v+u+\xi-\omega_p) \:
e^{i\omega_p(\tau_s-\tau_i)} \bigg| ^2 \: .
\end{eqnarray}

Equations \ref{IQ} and \ref{IC} can be calculated numerically by
assigning the appropriate $f(\omega,\Omega_0)$, $g(\xi)$, the
spectral amplitude of the pump $\textsf{A}_p(\omega_p)$ and the
power spectrum of the down-converted light $n_{s,i}(\omega)$.
However, these expressions can be further simplified analytically
by making a few more reasonable assumptions, according to the
specific nonlinear interaction that is being evaluated. Our only
assumption so far was that the pump bandwidth $\delta_p$ is
significantly smaller than the down-converted spectrum
$\Delta_\textsc{dc}$. This assumption enabled us to neglect the
dependence of $\Delta k$ on spectrum of the pump, and replace
$\omega_p$, with its center value $\langle \omega_p \rangle$.
Similarly, we will assume that $n(\omega)$ and $f(\omega)$ are
also roughly constant within spectral slices that are narrower
than the pump bandwidth $\delta_p$. In order to simplify Eqs.
(\ref{IQ}) and (\ref{IC}), we will assume from here on that the
bandwidth $\gamma$ of the final-state $g(\xi)$ is also
significantly smaller than the down-converted bandwidth:
\begin{eqnarray} \label{g<<D}
\Delta_\textsc{dc} \gg \delta_p  , \: \gamma \ .
\end{eqnarray}

In TPA $\gamma$ represents the bandwidth of the final level
$\gamma_f$, and in SFG it represents the possible (phase-matched)
bandwidth for the up-conversion process. In accordance with this
assumption we will neglect the dependence of $n_{s,i}$ on $\xi$,
which leads to:

\begin{eqnarray}
\label{IqcGeneral} I^c(t,\tau_s,\tau_i) &=&  \frac{1}{2\pi I_p} \:
\bigg| \int_\gamma d\xi \: g(\xi) \:  \textsf{A}_p(\Omega_0+\xi)
\: e^{-i\xi (t-\tau_i)}\: \bigg|^2 \nonumber \\ &\times&  \bigg|
\int_0^{\infty} f(\omega,\Omega_0) e^{-i\Delta k (\omega) L}
\sqrt{\left(1+n_s(\omega) \right)n_i(\Omega_0 -\omega)} \:
e^{-i\omega(\tau_i-\tau_s)} \:
d\omega \: \bigg|^2 \nonumber \\[5mm] I^{ic}(t,\tau_s,\tau_i) &=&
\frac{1}{(2\pi I_p)^2} \int_0^\infty du \: \bigg| \:
f(\langle\omega_p\rangle-u,\Omega_0) \: \sqrt{n_s(\langle \omega_p
\rangle-u) n_i(\Omega_0-\langle \omega_p \rangle +u)} \: \bigg|^2
\nonumber \\ &\times& \int_0^\infty dv \: \bigg| \int_\gamma d\xi
\ g(\xi)\:e^{-i\xi(t-\tau_i)} \int_0^\infty d\omega_p \:
\textsf{A}_p(\omega_p) \: \textsf{A}_p(v+u+\xi-\omega_p) \:
e^{i\omega_p(\tau_s-\tau_i)}
\bigg| ^2 \nonumber \\
\end{eqnarray}

To further clarify these expressions, let us further restrict
ourselves only to SFG and non-resonant TPA, for both of which we
can approximate the amplitude of spectral function
$f(\omega,\Omega_0)$ to be some average value $f_0$ over some
spectral bandwidth, $\Delta$, when $\Delta$ is taken to denote
only the part of this spectrum that overlaps with the
down-converted spectrum $\Delta_\textsc{dc}$. We will not neglect,
however, the \textit{phase} of $f(\omega,\Omega_0)$; specifically
- let us assume that the signal and the idler have spectral phases
of $\theta_s(\omega_s),\theta_i(\omega_i)$, respectively (for
example, due to spectral filters or a pulse-shaper). Thus we take
$f(\omega,\Omega_0)$ to be:
\begin{eqnarray} \label{theta s,i}
f(\omega,\Omega_0) &\approx& \ f_{avg} \: \times \:
e^{i[\theta_s(\omega)+\theta_i(\Omega_0-\omega)]}  \ \ when \
\omega \ lies \ within \ the \ bandwidth \ \Delta \nonumber \\
&\approx& \ 0 \ \ \ \ \ \ \ \ \ \ otherwise \:,
\end{eqnarray}
\noindent with $\Delta\leq\Delta_\textsc{dc}$. Let us also focus
our attention from here on to two-photon interactions with a
final-state frequency that is close to the pump frequency (as will
soon become evident, outside this regime the coherent signal dies
out, leaving only the incoherent signal):
\begin{eqnarray}
\label{O-Wp<D} |\Omega_0-\langle \omega_p \rangle | \ll \Delta \ .
\end{eqnarray}
We now define an "effective pulse" $P_e(t)$ which is the power of
a pulse with a power-spectrum that is equal to :
$n_e(\omega)=\big(n_s(\omega)+1\big)n_i(\Omega_0-\omega)$, and
spectral phase that is equal to
$e^{i[\theta_s(\omega)+\theta_i(\Omega_0-\omega)]}$:
\begin{equation}\label{Peff}
    P_e(t)=\frac{2\pi}{P_0} \ \bigg| \: \emph{\large F}^{\:-1}\bigg\{ \sqrt{\left(1+n_s(\omega)
\right)n_i(\Omega_0 -\omega)} \:
e^{i[\theta_s(\omega)+\theta_i(\Omega_0-\omega)]} e^{-i\Delta k
(\omega) L }\bigg\} \: \bigg|^2 \ ,
\end{equation}
\noindent where the factor of $2\pi$ originates from the symmetric
definition of the Fourier transform as : \\
\mbox{$x(t)=\emph{F}^{-1}\big\{\textsc{x}(\omega)\big\}=\frac{1}{\sqrt{2\pi}}\int
X(\omega)\:e^{-i\omega t}d\omega$}, and with $P_0$ being the total
power (times $2\pi$) of such a pulse with a constant spectral
phase (a 'transform-limited' pulse):
\begin{eqnarray}\label{P0}
    P_0=\bigg| \: \int_\Delta \sqrt{\big(n_s(\omega)+1\big)n_i(\Omega_0-\omega)} \:
d\omega \: \bigg|^2 \approx \Delta^2\: (n^2+n) \ ,
\end{eqnarray}

\noindent where $n=\langle n_s(\omega) \rangle \approx \langle
n_i(\Omega_0-\omega) \rangle$ is $2\pi$ times the average of the
photon flux spectral density of the down-converted light over the
bandwidth $\Delta$ (see Eqs. (\ref{n}) - (\ref{2pnS})). Due to the
normalization by $P_0$ we obtain:
\begin{equation}\label{Pe01}
    0 \leq P_e(t) \leq 1 \ ,
\end{equation}
\noindent where $P_e(0)=1$ is achieved for the un-shaped
(transform-limited) pulse. Finally we also take not of the fact
that the signals depend only on $t-\tau_i$ and $ \tau_i-\tau_s$:
$I(t,\tau_s,\tau_i)=I(t-\tau_i,\tau_i-\tau_s)$.\\

Using the definitions of Eqs. (\ref{P0}), (\ref{Peff}),
$I^c(t,\tau_s,\tau_i)$, $I^{ic}(t,\tau_s,\tau_i)$ take the
following form:

\begin{eqnarray}
\label{IqcEff} I^c(t-\tau_i,\tau_i-\tau_s) &\approx&  \big|
f_{avg} \big|^2 \: \frac{\Delta^2\: (n^2+n)}{2\pi I_p} \:   \bigg|
\int_\gamma d\xi \: g(\xi) \: \textsf{A}_p(\Omega_0+\xi) \:
e^{-i\xi (t-\tau_i)}\: \bigg|^2 \: P_e(\tau_i-\tau_s) \nonumber \\
[5mm] I^{ic}(t-\tau_i,\tau_i-\tau_s) &\approx& \big| f_{avg}
\big|^2 \: \frac{n^2}{(2\pi I_p)^2} \ \int_\Delta du \int_0^\infty
dv \nonumber \\
&\times&  \bigg| \int_\gamma d\xi \ g(\xi)\:e^{-i\xi(t-\tau_i)}
\int_0^\infty d\omega_p \: \textsf{A}_p(\omega_p) \:
\textsf{A}_p(v+u+\xi-\omega_p) \: e^{-i\omega_p(\tau_i-\tau_s)}
\bigg| ^2 \ .
\end{eqnarray}

\section{The results: the temporal and spectral behaviors of the
coherent vs. incoherent signals} \label{results}

\subsection{General}

\begin{figure} [t]
\begin{center}
\includegraphics[width=8.6cm] {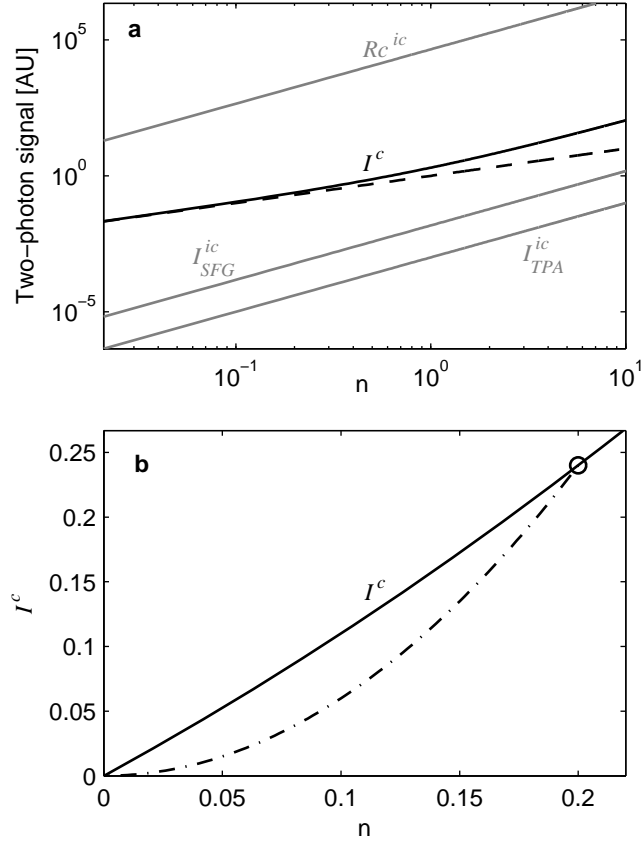}
\caption{\label{Fig LinSFG} Power-dependence of the coherent and
incoherent parts of two-photon interactions with down-converted
light. (a) The relative magnitudes of the coherent signal $I^c$
(black line) vs. the incoherent signals for TPA, SFG and
coincidence events ($I^{ic}_\textsc{tpa}$, $I^{ic}_\textsc{sfg}$
and $Rc^{ic}$, respectively, gray lines), assuming typical
physical parameters described in the text, represented on a
log-log scale. The graph shows the $n^2+n$ dependence of $I^c$,
with $n$ being the average spectral photon density, i.e. the total
photon flux in each of the signal and idler modes, divided by
their bandwidth $\Delta$. All the incoherent signals demonstrate a
quadratic dependence on $n$. The dashed line represents a
completely linear function, for comparison. (b) The dependence of
$I^c$ on $n$ represented on a linear scale, demonstrating the
nearly linear dependence at small values of $n$. The dash-dot line
depicts the quadratic response of $I^c$ assuming that
down-converted light with $n=0.2$ is being attenuated by linear
losses. }
\end{center}
\end{figure}

\begin{figure}[t!]
\begin{center}
\includegraphics[width=17cm] {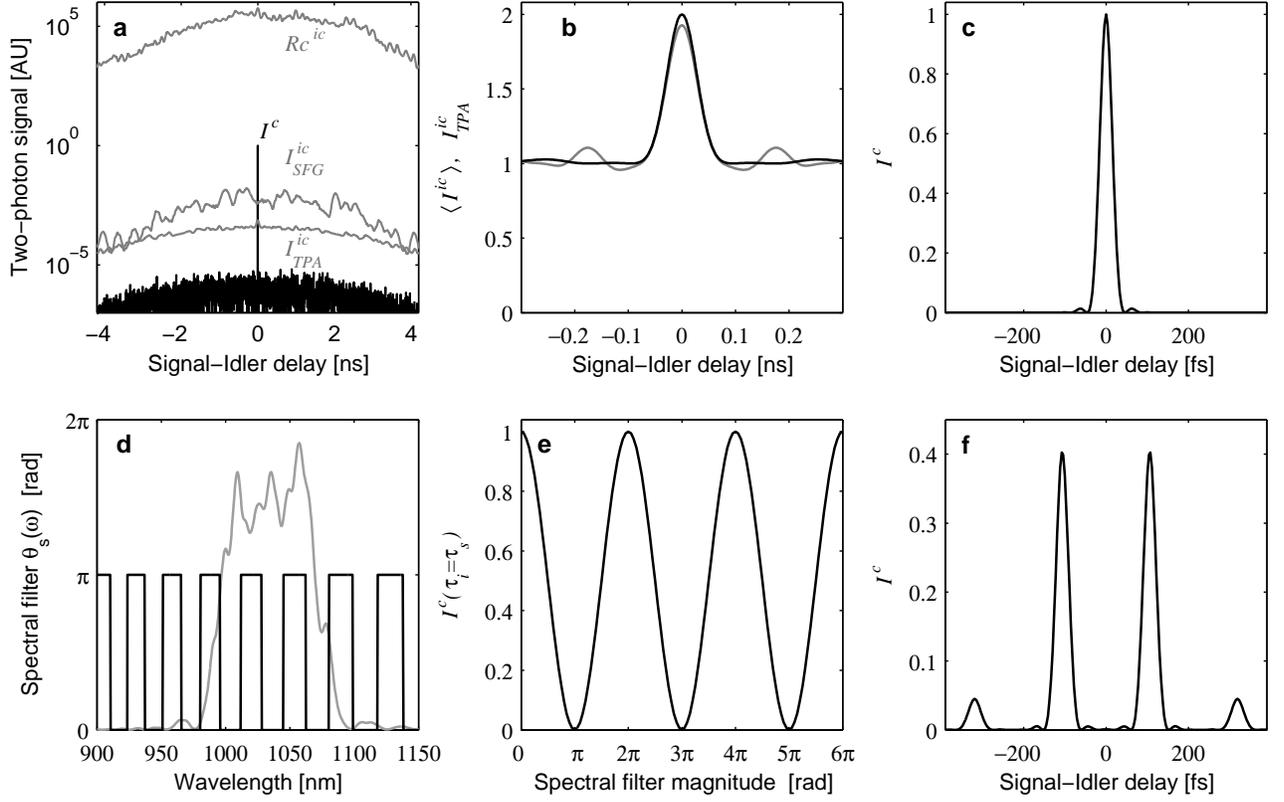}
\caption{\label{Fig temporal} The temporal behavior of the
coherent and incoherent parts of two-photon interactions with
down-converted light, as a function of a relative delay between
the signal and the idler fields. (a) The relative magnitudes of
the coherent signal $I^c$ (black line) vs. the incoherent signals
for TPA, SFG and coincidence events ($I^{ic}_\textsc{tpa}$,
$I^{ic}_\textsc{sfg}$ and $Rc^{ic}$, respectively, gray lines),
assuming typical physical parameters (described in the text),
represented on a logarithmic scale. This graph depicts the
instantaneous peak of the signals at $t=\tau_i$ for one typical
down-converted pulse, and shows the contrast between the sharp
temporal behavior of $I^c$, and the behavior of the incoherent
signals, which is on the same ns-timescale as the pump pulse. Note
that $I^{ic}_\textsc{tpa}$ presents a smooth and symmetric graph,
reflecting fact that the long-lived ($\sim30\:$ns) final atomic
state integrates over the intensity variations of the pump pulse.
Similarly, $Rc^c$, which takes into account integration over the
gating time $Tg\sim 1\:$ns is more smooth than $I^c_\textsc{sfg}$,
which represents an instantaneous parametric process. For the case
assumed here of a non transform-limited $3\:$ns pump pulse, the
ensemble average of all the incoherent signals becomes
proportional to the normalized intensity-correlation function of
the pump $\texttt{g}^{(2)}_p(\tau_i-\tau_s)$, which exhibits a
$\sim \times 2$ 'bunching peak' at delays which are shorter than
the pump's coherence time $2\pi/\delta_p$. This is shown in (b) in
black, together with the instantaneous incoherent TPA signal
$I^{ic}_\textsc{tpa}$ (gray), which can be approximated to be
proportional to $\texttt{g}^{(2)}_p$ even without averaging.  (c)
A zoomed-in presentation of the ultrashort-pulse like behavior of
the coherent signal: $I^c(\tau_i-\tau_s)\propto
P_e(\tau_i-\tau_s)$, assuming dispersion is compensated and no
additional spectral filtering, resulting in a behavior that is
identical to that of mixing two $35\:$fs transform-limited pulses.
(e) $I^c$ at zero signal-idler delay as a function of the
magnitude of a spectral phase filter $\theta_s(\omega)$ that is
applied to the signal, for example by a pulse-shaper.
$\theta_s(\omega)$ is drawn in (d) (black line) at a magnitude of
$\pi$, together with the down-converted power spectrum (gray
line). (f) The shaped temporal behavior of the coherent signal
with $\theta_s(\omega)$ as depicted in (d) applied to the signal
field.}
\end{center}
\end{figure}

\begin{figure}[t!]
\begin{center}
\includegraphics[width=8.6cm] {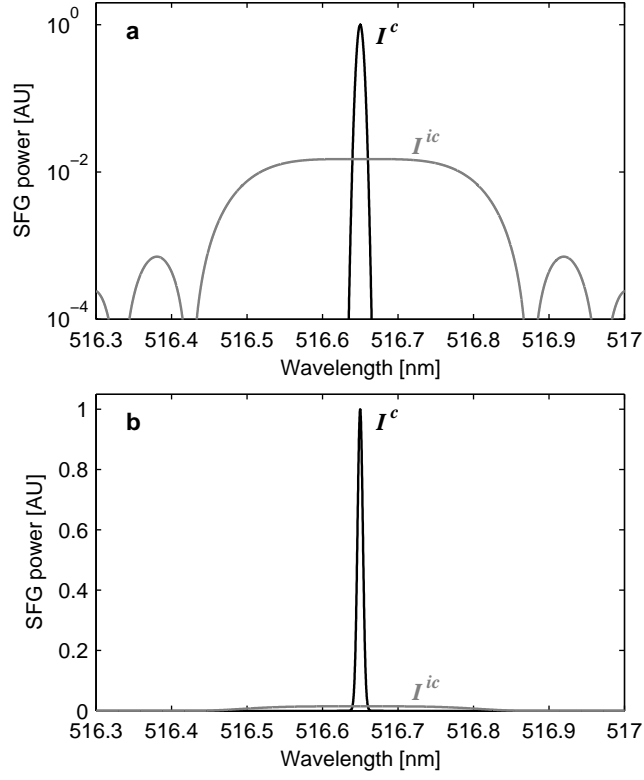}
\caption{\label{Fig SFG spectrum} The spectrum of the coherent and
incoherent parts of SFG light induced by high-power ($n \gg 1$)
broadband down-converted light, with physical parameters as
described in the text. The graphs are shown on a logarithmic scale
in (a) and a linear scale in (b), and demonstrate how the coherent
signal (black) replicates the narrow spectrum of the pump
($\delta_p=0.01nm$). In contrast, the spectrum of the incoherent
signal (gray) is as wide as the phase-matching conditions at the
crystal allow it to be ($\gamma_\textsc{uc}=0.3nm$).}
\end{center}
\end{figure}

\begin{figure}[t!]
\begin{center}
\includegraphics[width=8.6cm] {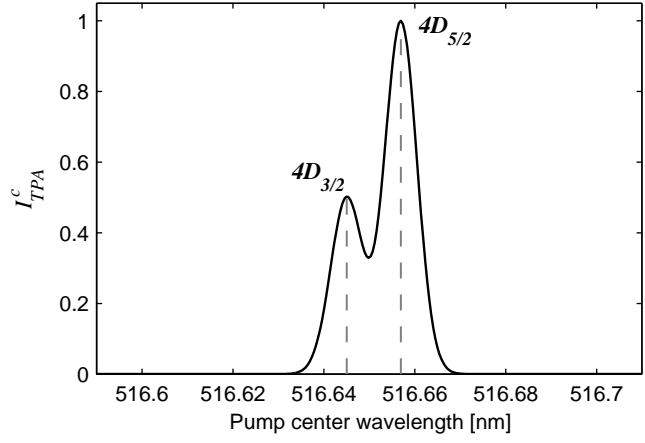}
\caption{\label{Fig TPA spectrum} The dependence of the coherent
part of TPA ($I^c_\textsc{tpa}$, black) induced by down-converted
light, as a function of the center frequency of the pump. The
graph shows how $I^c_\textsc{tpa}$ behaves as if the $0.01nm$-wide
pump itself is inducing the transition ($5S_{1/2}$ $\rightarrow$
$4D_{3/2}$,$4D_{5/2}$ in atomic Rb, with the transition
wavelengths drawn in gray). The incoherent part of the TPA is
practically insensitive to the pump wavelength, and since for
practical power levels ($n>1$) it is smaller than the coherent
part by roughly $\delta_p/\Delta\approx 1/2000$ (with $\delta_p$,
$\Delta$ being the bandwidths of the pump and the down-converted
light, respectively), it is too small to appear in this graph.}
\end{center}
\end{figure}

Equation \ref{IqcEff} already reveals most of the unique features
of two-photon interactions induced by broadband down-converted
light that were mention at the introduction. These features are
presented in Figs. \ref{Fig LinSFG}-\ref{Fig TPA spectrum}. In
order to present a quantitative picture of the behavior and the
relative magnitudes of the coherent and incoherent signals, we
assume in the following realistic physical parameters, that are
similar to the experimental parameters in
\cite{{Dayan@Silberberg_QPH_2003},Dayan@Silberberg_PRL_2004}.
Specifically, we assume broadband, degenerate but non-collinear
down-converted light at a bandwidth of $80nm$ around $1033nm$, and
consider two-photon interactions around $\Omega_0=516.5 \:$nm.
Specifically, for TPA we assume a final-level bandwidth of
$\gamma_\textsc{tpa}\approx5MHz$, and for SFG we assume a
phase-matched bandwidth for up-conversion of
$\gamma_\textsc{uc}=0.3nm$. We assume the down-converted light was
generated by a pump that is a $\sim 3\:$ns pulse with a bandwidth
of $\delta_p=0.01nm$ around $516.5nm$. Such parameters are typical
for Q-switched laser systems.\\

The first unique feature is the coherent signal's non-classical
linear intensity dependence: $I^c \propto n^2 +n$. This behavior
manifests the fact that the signal and idler modes share a
\textit{single} wavefunction. Figure \ref{Fig LinSFG} depicts this
behavior of the coherent signal, compared to the quadratic
intensity dependence of the incoherent signal. The relative
magnitudes of the incoherent signals for SFG, TPA and coincidence
events are calculated using Eqs. (\ref{ratio SFG}),
(\ref{ratioTPA})
and (\ref{ratio Rc}), respectively (derived in the following subsections). \\

Note that while the dependence of the coherent signal on the flux
of the down-converted photons may be linear, the response of the
two-photon signal (TPA, SFG or coincidences) to
\textit{attenuation} of the down-converted light by linear losses
(namely absorption or scattering, for example by optical filters
or beam splitters) is always quadratic, as is evident from the
presence of the term $\big| f_{avg} \big|^2$ in the expressions
for both the coherent and incoherent signals. This behavior is
depicted by the dash-dot line in Fig. \ref{Fig LinSFG}, which
assumes that down-converted light with average spectral photon
density of $n=0.2$ is being attenuated by optical filters. These
results are in excellent agreement with the experimental results
of SFG with entangled photons presented in
\cite{Dayan@Silberberg_PRL_2005}.\\

The second unique feature that appears in Eq. (\ref{IqcEff}) is
the pulse-like response of the coherent interaction to a relative
delay between the signal and the idler, as represented by the term
$P_e(\tau_i-\tau_s)$, which is the (normalized) response of mixing
two ultrashort pulses with practically the same spectra as the
signal and idler. As such, $P_e(\tau_i-\tau_s)$) is sensitive to
dispersion, including the dispersion that was accumulated in the
down-conversion process itself, denoted by the term $e^{-i\Delta k
(\omega) L}$ in the definition of $P_e(\tau_i-\tau_s)$ (Eq.
(\ref{Peff})). Figure \ref{Fig temporal}(c) shows a zoomed-in
picture of this sharp temporal dependence of the coherent signal
on relative delay between the signal and the idler beams, assuming
that dispersion is either negligent or is compensated by spectral
phase filters, leading to a sharp response which is exactly as if
the interaction was induced by a pair of $35fs$
(transform-limited) pulses.\\

The coherent summation over the spectrum which leads to this
pulse-like behavior also implies that the coherent signal can be
shaped by spectral-phase manipulations, exactly like with coherent
ultrashort pulses. Note that the shape of $P_e(\tau_i-\tau_s)$ is
determined by the sum of the phases applied to antisymmetric
spectral components of the signal and the idler:
$\theta_s(\omega)+\theta_i(\Omega_0-\omega)$. This implies that if
the same phase filter is applied to both the signal and the idler
beams (or the same dispersive medium), only spectral phase
functions that are symmetric about $\Omega_0/2\approx\big\langle
\omega_p \big\rangle /2$ would affect $P_e(\tau_i-\tau_s)$.
Figures \ref{Fig temporal}(c)-(e) demonstrate how applying a
spectral phase filter to the signal (or idler) spectrum leads to
the same result as with coherent ultrashort pulses. This
ultrashort-pulse-like behavior (including the ability to tailor it
by a pulse-shaper) was demonstrated experimentally with high-power
SFG in \cite{Dayan@Silberberg_QPH_2003}, high power TPA in
\cite{Dayan@Silberberg_PRL_2004}, and with broadband entangled
photons in \cite{Peer@Silberberg_PRL_2005}, with excellent
agreement with our calculations.\\

Interestingly, $P_e(\tau_i-\tau_s)$, \textit{does not} depend on
the specific type of the two-photon interaction, i.e. the coherent
signal of TPA, SFG or coincidence events will always exhibit this
ultrashort-pulse like behavior. The contrast between the temporal
behavior of the coherent signal and that of the incoherent ones is
shown in Fig. \ref{Fig temporal}(a). Note that Fig. \ref{Fig
temporal}(a) presents the instantaneous peaks of the coherent and
incoherent signals at $t=\tau_i$ for one, single-shot arbitrary
example. As is evident, the incoherent signals (calculated for
TPA, SFG and coincidence events by Eqs. (\ref{Iic narrow pump}),
(\ref{IcTPA}) and (\ref{Rcic Tg}), respectively, assuming $n \gg
1$) always demonstrate a temporal dependence on $\tau_s-\tau_i$
that is on the same $ns$-timescale as the pump pulse.\\

It is important to note that the duration of such Q-switched
pulses is much longer than their coherence time (which in this
case is $2\pi/\delta_p\approx 89\:$ps), which was their duration
if they were transform-limited. Such pump pulses, for which:

\begin{eqnarray} \label{quasicw}
\tau_p \gg 2\pi/\delta_p \ ,
\end{eqnarray}

\noindent can be considered a 'quasi-continuous' light, since they
can be viewed as short bursts of continuous light, especially when
time-scales that are shorter than $\tau_p$ are considered, during
which the average intensity stays roughly constant. Thus, such
quasi-continuous pump pulses yield approximately the same results
for $I^c$, $I^{ic}$ as a continuous pump, especially when the
ensemble average of many such pulses is considered. In particular,
as will be shown in the next subsections, once averaged the
incoherent signals all becomes proportional to the normalized
second-order correlation function of the \textit{pump}
$\texttt{g}^{(2)}_p(\tau_i-\tau_s)$. This behavior is depicted in
Fig. \ref{Fig temporal}(b), which depicts the calculated
$\texttt{g}^{(2)}_p(\tau_i-\tau_s)$ together with a zoomed-in
presentation of the instantaneous incoherent TPA signal
$I^{ic}_\textsc{tpa}$. As is evident, even without averaging
$I^{ic}_\textsc{tpa}$ follows very closely
$\texttt{g}^{(2)}_p(\tau_i-\tau_s)$. This is explained by the fact
that the incoherent excitation of the long-lived ($\sim 30 \:$ns)
final atomic state actually averages the intensity fluctuations of
the pump (see Eq. (\ref{IcTPA})). When the ensemble average of
many such quasi-continuous pulses is taken, the incoherent TPA
signal, as well as $I^{ic}_\textsc{sfg}$ and $Rc^{ic}$ become
practically identical to $\texttt{g}^{(2)}_p(\tau_i-\tau_s)$,
demonstrating the expected $\times 2$ 'bunching-peak' at delays
which are shorter than the coherence length of the pump
$(\tau_s-\tau_i <
1/\delta_p)$ \cite{Mandel&Wolf_1995}.\\

Another result of the temporal integration which is performed by
the incoherent TPA process, is the fact that the behavior of
$I^{ic}_\textsc{tpa}(\tau_i-\tau_s)$ is more smooth and symmetric
than that of $I^{ic}_\textsc{sfg}(\tau_i-\tau_s)$, which
represents an instantaneous process. Since coincidence measurement
also includes a temporal integration over the $\sim 1\:$ns
gating-time of the detectors, $Rc^{ic}$ presents a behavior which
is more smooth and symmetric than $I^{ic}_\textsc{sfg}$, but not
as much as $I^{ic}_\textsc{tpa}$.\\

Unlike the incoherent signals, the coherent signal's sharp
behavior $P_e(\tau_i-\tau_s)$ depends on the large-scale
properties of the down-converted spectrum, and therefore its shape
is not affected by shot-to-shot noise, only its relative height.
Thus, we see that there are three timescales in our system. One is
the duration of the pump pulse (which can be infinity for a
continuous pump), the other is the coherence time of the pump
which is $\sim 1/\delta_p$ (and is equal to the duration of the
pump pulse, in case it is a transform-limited one), and the
shortest time scale is the behavior of the coherent signal, which
is on the same timescale as the coherence time of the
down-converted light: $1/\Delta$. In the case considered in Fig.
\ref{Fig temporal}, the $35fs$ pulse-like behavior of the coherent
signal stands in contrast with the temporal behavior of the
down-converted light itself, which is a $3ns$ pulse in this case,
i.e. 85,000 times longer. The effect is of course even more
intriguing when continuously-pumped down-conversion is
considered.\\

The temporally-sharp behavior of the coherent signal also stands
in contrast with its equally sharp spectral behavior. More
specifically, we need to distinguish between two spectral
behaviors. One is the excitation-spectrum, i.e. the frequencies
which are excited by the two-photon interaction. The other is the
dependence of the interaction on the pump wavelength. In the case
of SFG, the excitation spectrum corresponds to the spectrum of the
up-converted light. In the case of TPA this spectrum corresponds
to which atomic levels will be excited. By Fourier-transforming
the amplitude of the coherent signal in Eq. (\ref{IqcEff}) back to
the frequency domain $\Omega$ of the generated signal, we see
immediately that the excitation power-spectrum of the coherent
signal is simply the spectral overlap between the narrowband pump
and the final state, and does not reflect the broad spectra of the
signal and the idler fields which induce the interaction:

\begin{eqnarray}
\label{IqW} I^c(\Omega,\tau_s,\tau_i) &\approx&  \big| f_{avg}
\big|^2 \: \frac{\Delta^2\: (n^2+n)}{I_p} \: P_e(\tau_i-\tau_s) \:
\bigg| g(\Omega-\Omega_0) \: \textsf{A}_p(\Omega) \:
e^{i\xi\tau_i} \bigg|^2  \: .
\end{eqnarray}

This implies that if the pump bandwidth is narrower than the final
state bandwidth, the excitation spectrum $I^c(\Omega)$ would
follow that of the narrowband pump, as is shown for SFG in Fig.
\ref{Fig SFG spectrum}. In other words, the coherent signal
behaves as if the pump itself was inducing the interaction. While
the spectral behavior of the incoherent signal is harder to deduce
out of Eq. (\ref{IqcEff}), in the following we show that it is
approximately that of the final state; this is shown more easily
if we assume the final state is significantly broader than the
pump (Eq. (\ref{Iic narrow pump})), or the other way around (Eq.
(\ref{Iic broad pump})). This spectral behavior of the coherent
and incoherent was demonstrated experimentally in
\cite{Abram@Dolique_PRL_1986,{Dayan@Silberberg_QPH_2003}}.\\

For the case of TPA, even if the pump is narrower than the final
atomic state, this is not reflected in the spectrum of the
fluorescence from that level, because the temporally random,
incoherent emission process of the emission erases the information
on the exact frequency that drove the transition (especially in
this limit of weak, non-stimulated interaction). However, while
the excitation spectrum may not be directly accessible, the other
kind of spectral behavior, i.e. the dependence of TPA on the pump
wavelength, can be explored experimentally. As already evident
from Eq. (\ref{IqcEff}), the magnitude of the coherent signal
depends on the total spectral overlap between the pump spectrum
and the final state spectrum. This is shown in Fig. \ref{Fig TPA
spectrum}, which considers TPA in atomic Rubidium (Rb) at the
$5S_{1/2}-4D_{3/2,5/2}$ transition. The coherent TPA rate indeed
behaves as if the pump laser itself was inducing the transition
(which is of course a forbidden transition for a one-photon
process), demonstrating a spectral resolution of 0.01nm, almost
resolving the $13.4GHz$ hyperfine splitting between the $4D_{3/2}$
and the $4D_{5/2}$ levels, even though the interaction is induced
by a light with a total bandwidth that is $\sim 2000$ times wider.
For the sake of simplicity we ignored in Fig. \ref{Fig TPA
spectrum} the hyperfine splitting of the $5S_{1/2}$ ground-level
in Rb ($3GHz$ for Rb87, $6.8GHz$ for Rb85). This calculation
should be compared with the experimental results of
\cite{Dayan@Silberberg_PRL_2004} (in which a wider pump bandwidth
of $0.04nm$ and power-broadening prevented resolving the
hyper-fine splitting, nonetheless demonstrating a spectral
resolution that was $\sim 2000$ times narrower than the
down-converted bandwidth). In contrast, the incoherent signal (for
SFG, TPA and of course coincidence events) is practically
independent of the wavelength of the pump (see Eq. (\ref{Iic
narrow pump}) for SFG and Eq. (\ref{Iic broad pump}) for TPA). The
incoherent signal responds only to the change in the
down-converted power spectrum that results from the change of the
pump wavelength, and so it exhibits only the very wide spectral
response that is expected in an interaction that is induced by
mixing of two $80nm$ wide incoherent beams. This can be viewed as
resulting from the fact that the incoherent interaction
\textit{has no knowledge} of what was the wavelength of the pump
that generated the down converted light, since that "information"
lies only in the correlations between the spectral phases of the
down-converted modes - phases that play no role in the generation
of the incoherent signal.\\

In the following subsections we present more specific analytic
expressions for the temporal behavior of the coherent and
incoherent signals, performing approximations that fit various
two-photon interactions. First we consider the case where the
spectral width of the final state exceeds that of the pump, as is
typically the case with SFG and is always the case with
coincidence detection. Then we consider the case where the
spectral width of the final state is smaller than that of the
pump, as can occur with TPA. For more accurate results, and for
experimental schemes that do not comply completely with the
assumptions and approximations that
follow, Eqs. (\ref{IqcGeneral}) or (\ref{IqcEff}) should be used.\\

\subsection{Pump bandwidth smaller than the final-state bandwidth (example: SFG)}

So far we have assumed that the down-converted bandwidth is
significantly larger than the pump bandwidth and the final-state
bandwidth (Eq. (\ref{g<<D})). In the following derivation we
consider the case where the pump bandwidth is also significantly
narrower than the final-state bandwidth:

\begin{eqnarray} \label{dp<<g}
\Delta  \gg \gamma \gg \delta_p \ .
\end{eqnarray}

This condition allows us to simplify the expressions for both the
coherent and the incoherent signals from Eq. (\ref{IqcEff}) by
using the following approximation:

\begin{eqnarray}
\label{narrow pump approx} \int_\gamma d\xi \: g(\xi) \:
\textsf{A}_p(\bar{\omega}+\xi) \: e^{-i\xi (t-\tau_i)} &\approx&
\sqrt{2\pi} \: g(\langle \omega_p \rangle-\bar{\omega}) \:
A_p(t-\tau_i) \: e^{i\bar{\omega}(t-\tau_i)} \ ,
\end{eqnarray}

\noindent where
$A_p(t)=\emph{F}^{-1}\big\{\textsf{A}_p(\omega)\big\}$ is the
temporal amplitude of the pump, leading quite immediately to:

\begin{eqnarray}
\label{Ic narrow pump} I^c(t-\tau_i,\tau_i-\tau_s) &\approx& \big|
g(\langle \omega_p \rangle - \Omega_0)  \big|^2 \: \big| f_{avg}
\big|^2 \:
 \frac{\texttt{I}_p(t-\tau_i)}{I_p} \:
\Delta^2\: (n^2+n) \: P_e(\tau_i-\tau_s) \ .
\\ \label{Iic narrow pump}
I^{ic}(t-\tau_i,\tau_i-\tau_s) &\approx& \ \int_\gamma d\xi' \big|
g(\xi') \big|^2 \: \ \big| f_{avg} \big|^2 \:
\frac{\texttt{I}_p(t-\tau_i) \texttt{I}_p(t-\tau_s)}{I_p^2} \:
\Delta \: n^2 \ ,
\end{eqnarray}

\noindent with $\texttt{I}_p(t)=|A_p(t)|^2$. To represent more
accurately typical experimental conditions, we can take the
ensemble average of the signals (i.e. averaging over many pulses
in the case of a pulsed pump, or over time in the case of a
continuous pump). In particular, if we consider a continuous pump
or a quasi-continuous pump (i.e. the center part of a
non-transform-limited pump pulse for which $\tau_p\gg
2\pi/\delta_p$), we can approximate for $t-\tau_s,t-\tau_i \ll
\tau_p$:

\begin{eqnarray} \label{Ipt=Ip0}
\big\langle \texttt{I}_p(t-\tau_i) \big\rangle \approx \big\langle
\texttt{I}_p(t-\tau_i) \big\rangle \approx \big\langle
\texttt{I}_p(0) \big\rangle \approx I_p \ .
\end{eqnarray}

\noindent This allows us to further simplify the expressions for
the coherent and incoherent signals by using:

\begin{eqnarray}
\label{Ip0/Ip=1}
\frac{\big\langle\texttt{I}_p(t-\tau_i)\big\rangle}{I_p} &\approx&
1 \ ,
\\ \label{IptIpt=g2}
\frac{\big\langle \texttt{I}_p(t-\tau_i)
\texttt{I}_p(t-\tau_s)\big\rangle}{I_p^2} &\approx&
\frac{\big\langle \texttt{I}_p(t-\tau_i)
\texttt{I}_p(t-\tau_s)\big\rangle}{\big\langle
\texttt{I}_p(t-\tau_i) \big\rangle \big\langle
\texttt{I}_p(t-\tau_s)\big\rangle} = \texttt{g}^2_p(\tau_i-\tau_s)
\ ,
\end{eqnarray}

\noindent where $\texttt{g}^2_p(\tau_i-\tau_s)$ is the normalized
second-order correlation function of the pump field. Note that the
pump field was taken as a classical amplitude throughout this
paper, hence $\texttt{g}^2_p$ represents in our derivation only
classical intensity correlations. This leads to:

\begin{eqnarray}
\label{Ic narrow pump av} \big \langle I^c(t-\tau_i,\tau_i-\tau_s)
\big \rangle &\approx& \big| g(\langle \omega_p \rangle -
\Omega_0) \big|^2 \: \big| f_{avg} \big|^2 \:
 \Delta^2\: (n^2+n) \: P_e(\tau_i-\tau_s) \ .
\\ \label{Iic narrow pump av}
\big \langle I^{ic}(t-\tau_i,\tau_i-\tau_s) \big \rangle &\approx&
\ \int_\gamma d\xi' \big| g(\xi') \big|^2 \: \ \big| f_{avg}
\big|^2 \: \texttt{g}^2_p(\tau_i-\tau_s) \: \Delta \: n^2 \ .
\end{eqnarray}

As is reflected in the term $\int_\gamma d\xi' \big| g(\xi')
\big|^2$ in Eqs. (\ref{Iic narrow pump}), (\ref{Iic narrow pump
av}), the power of the incoherent signal is indeed the incoherent
summation of all its spectral components, and its spectrum follows
that of the final state. This behavior is depicted in Fig.
\ref{Fig SFG spectrum} for SFG, in which case we may write:

\begin{eqnarray} \label{kappa SFG}
\big| f_{avg} \big|^2 \ \big| g(\Omega-\Omega_0) \big|^2 = L^2 \:
\textrm{sinc}^2\big[ \Delta k( \langle \omega \rangle, \Omega)
L/2\big] \: \beta^2(\langle \omega\rangle,\Omega) \ .
\end{eqnarray}

With the exception of very long crystals (i.e. more than $\sim 1$
cm in the case of type-I up conversion, or more than a few mm in
the case of type-II up conversion), the up-converted bandwidth
$\gamma_\textsc{uc}$ is typically at least of the order of $0.01
\:$THz, and could even reach tens of THz for very short crystals.
This bandwidth is orders of magnitude larger than the typical
bandwidth of continuous lasers, and even larger than the bandwidth
of Q-switched, $ns$-pulsed lasers. Thus, the condition of Eq.
(\ref{dp<<g}) is typically satisfied for SFG,
unless ultrashort pulses (hundreds of $fs$ or less) are used.\\

Equations \ref{Iic narrow pump},\ref{Iic narrow pump av} displays
both the spectral and the temporal behaviors of the incoherent
signal more clearly than Eq. (\ref{IqcEff}) did. As is evident,
the incoherent signal is practically insensitive to the pump
wavelength, and its response to a relative delay between the
signal and the idler is very slow, since it depends on the
temporal behavior of the pump, which is either continuous or a
very long pulse, compared to the ultrashort-pulse-like behavior of
the coherent signal; as demonstrated by the term
$\texttt{I}_p(t-\tau_i) \texttt{I}_p(t-\tau_s)$, the incoherent
signal depends only on the temporal overlap between the
intensities of the signal and the idler (which follow the
intensity of the pump). This behavior is depicted in Fig. \ref{Fig
temporal}(a)-(c). In particular, Fig. \ref{Fig temporal}(b) shows
the 'bunching' peak which is exhibited by the incoherent signal
for a quasi-continuous pump at delays which are shorter
than the pump's coherence length. \\

Since the observed signal is always the sum of the coherent and
incoherent contributions, it interesting to compare the magnitude
of the coherent signal to that of the incoherent one. The average
ratio between the coherent signal and the incoherent one is
therefore:
\begin{eqnarray} \label{ratio SFG}
\frac{\big\langle I^{c}(t-\tau_i,\tau_i-\tau_s)\big\rangle}
{\big\langle I^{ic}(t-\tau_i,\tau_i-\tau_s)\big\rangle} &\approx&
\frac{\big\langle \texttt{I}_p(t-\tau_i) \big\rangle \:
I_p}{\big\langle \texttt{I}_p(t-\tau_i)
\texttt{I}_p(t-\tau_s)\big\rangle} \frac{\Delta}{ \gamma } \:
\frac{(n^2+n)}{n^2} \: P_e(\tau_i-\tau_s) \ ,
\end{eqnarray}

\noindent which for a continuous or quasi-continuous pump becomes:

\begin{eqnarray} \label{ratio SFG g}
\frac{\big\langle I^{c}(t-\tau_i,\tau_i-\tau_s)\big\rangle}
{\big\langle I^{ic}(t-\tau_i,\tau_i-\tau_s)\big\rangle} &\approx&
 \frac{\Delta}{ \gamma \: \texttt{g}^2_p(\tau_i-\tau_s)} \:
\frac{(n^2+n)}{n^2} \: P_e(\tau_i-\tau_s) \ .
\end{eqnarray}

In the absence of spectral phase filters, and assuming that
dispersion is corrected, we can assign $P_e(0)=1$. Thus, taking
into account that typically
$1\leq\texttt{g}^2_p(\tau_i-\tau_s)\leq2$, the expected ratio
between the coherent and incoherent signals approaches
$\Delta/\gamma$ for high powers ($n>1$), and even more for low
powers where $n+n^2 \gg n^2$. Thus, under the conditions of
broadband down conversion and a narrowband final state assumed
throughout our derivation, the coherent signal dominates over the
incoherent signals (see the relative magnitude of
$I^{ic}_\textsc{sfg}$ in Figs. \ref{Fig LinSFG}-\ref{Fig SFG
spectrum}).\\

\subsection{Pump bandwidth larger than the final-state bandwidth (example: TPA)}

For the following derivation we assume that the final-state
bandwidth $\gamma$ is significantly narrower than the pump
bandwidth, as may often be the case with TPA:

\begin{eqnarray} \label{g<<dp}
\Delta \gg \delta_p \gg \gamma \ .
\end{eqnarray}

Assuming this, we can simplify the expressions from Eq.
(\ref{IqcEff}) by assuming that the spectral amplitude of the pump
remains constant within spectral slices which are narrower than
$\gamma$ :

\begin{eqnarray}
\label{broad pump approx} \textsf{A}_p(\bar{\omega}+\xi) &\approx&
 \textsf{A}_p(\bar{\omega}) \ .
\end{eqnarray}

Additionally, in order to evaluate the incoherent TPA signal we
use Parseval's theorem to obtain the following relation:
\begin{eqnarray}
\label{u} \int^\infty_0 d\upsilon \: \bigg| \int_0^\infty d
\omega_p \: \textsf{A}_p(\omega_p) \:
\textsf{A}_p(\upsilon-\omega_p) e^{i\omega_p\tau} \bigg|^2 &=&
2\pi \int_{-\infty}^\infty \texttt{I}_p(t'+\tau)\texttt{I}_p(t')
\: dt' \ ,
\end{eqnarray}
\noindent leading to:

\begin{eqnarray}
\label{Ic broad pump} I^c(t-\tau_i,\tau_i-\tau_s) &\approx&  \big|
f_{avg} \big|^2 \: \big|\textsf{A}_p(\Omega_0)\big|^2 \:
\frac{\Delta^2\: (n^2+n)}{2\pi I_p} \:   \bigg| \int_\gamma d\xi
\: g(\xi) \:
e^{-i\xi (t-\tau_i)}\: \bigg|^2 \: P_e(\tau_i-\tau_s) \\
[5mm] \label{Iic broad pump} I^{ic}(t-\tau_i,\tau_i-\tau_s)
&\approx& \big| f_{avg} \big|^2 \: \frac{\Delta \: n^2}{2\pi
I_p^2} \ \int_{-\infty}^\infty
\texttt{I}_p(t'+\tau)\texttt{I}_p(t') \: dt' \: \  \bigg|
\int_\gamma d\xi \ g(\xi)\:e^{-i\xi(t-\tau_i)} \bigg| ^2 \ .
\end{eqnarray}

Equations \ref{Ic broad pump}-\ref{Iic broad pump} show that in
this case the spectra of both the coherent and the incoherent
signals are determined by the spectrum of the final state
$g(\xi)$.\\

To clarify the temporal behavior of both signals, let us use

\begin{eqnarray} \label{G(t)} \int_\gamma d\xi \
g(\xi)\:e^{-i\xi(t-\tau_i)}=G(t-\tau_i) \ ,
\end{eqnarray}
with $G(t)=\emph{F}^{-1}\big\{g(\xi)\big\}$ being the slowly
varying envelope of the temporal response of the final state. This
leads to:

\begin{eqnarray}
\label{Ic broad pump t} I^c(t-\tau_i,\tau_i-\tau_s) &\approx&
\big| f_{avg} \big|^2 \: \big|\textsf{A}_p(\Omega_0)\big|^2 \:
\frac{\Delta^2\: (n^2+n)}{I_p} \:   \big| G(t-\tau_i) \big|^2 \: P_e(\tau_i-\tau_s) \\
[5mm] \label{Iic broad pump t} I^{ic}(t-\tau_i,\tau_i-\tau_s)
&\approx& \big| f_{avg} \big|^2 \: \frac{\Delta \: n^2}{I_p^2} \
\int_{-\infty}^\infty \texttt{I}_p(t'+\tau)\texttt{I}_p(t') \: dt'
\: \  \big| G(t-\tau_i) \big| ^2 \ .
\end{eqnarray}

Considering TPA as a probable example for the case where the final
state is considerably narrower than the pump, we can substitute
$g(\xi)=\frac{1}{\xi + i\gamma}$. Using
$\emph{F}^{-1}\big\{g(\xi)\big\}=\textsc{u}(t-\tau_i) \:
e^{-\gamma_h(t-\tau_i)}$ (where $\textsc{u}(t)$ is a
step-function), we obtain for TPA:

\begin{eqnarray}
\label{IqTPA} I^c(t-\tau_i,\tau_i-\tau_s) &=& \kappa_\textsc{tpa}
\: \frac{\big|\textsf{A}_p(\Omega_0) \big|^2 }{I_p} \:
\textsc{u}(t-\tau_i) \: e^{-2\gamma_f(t-\tau_i)} \: \Delta^2\:
(n^2+n) \: P_e(\tau_i-\tau_s)  \\ \label{IcTPA}
I^{ic}(t-\tau_i,\tau_i-\tau_s) &=& \kappa_\textsc{tpa} \:
\frac{\int_{-\infty}^\infty
\texttt{I}_p(t'+\tau_s-\tau_i)\texttt{I}_p(t') \: dt'}{I_p^2} \:
\textsc{u}(t-\tau_i) \: e^{-2\gamma_f(t-\tau_i)} \: \Delta\: n^2 \
,
\end{eqnarray}

\noindent with
\begin{eqnarray} \label{kappa}
\kappa_\textsc{tpa} =  \frac{1}{32 \pi^3} \bigg[
\frac{\sum_{n}\mu_{fn}\mu_{ng} }{ c \varepsilon S \hbar}\bigg]^2
\frac{\langle \omega \rangle (\omega_{fg}-\langle \omega \rangle)
} {(\omega_{ng}-\langle \omega \rangle )^2} \ .
\end{eqnarray}

In order to have a better estimation of the magnitude of the
coherent TPA signal as compared to the incoherent one, let us try
to estimate the average magnitude of $\big|
\textsf{A}_p(\Omega_0)\big|^2$, assuming of course that $\Omega_0$
is within the pump spectrum (i.e. $|\Omega_0-\langle \omega_p
\rangle | < \delta_p/2$). For a pulsed pump, we can use Parseval's
theorem again to relate between the average spectral power of the
pump $\big\langle | \texttt{A}_p |^2 \big\rangle$ to the average
temporal power $I_p$:

\begin{eqnarray}\label{Apavg p}
  \delta_p \:  \langle \big| \textsf{A}_p\big|^2 \rangle &=&
    \tau_p  \: I_p \ .
\end{eqnarray}

However, to get a similar relation for a continuous pump we need
to identify the longest timescale in our system, which defines the
smallest frequency increment, i.e. the quantization unit of our
frequency domain. In the case considered in this subsection, the
smallest frequency scale is that of the final state, $\gamma_f$
(Eq. (\ref{g<<dp})), hence the longest relevant timescale is the
final state lifetime $\tau_f$. Accordingly we can approximate for
a continuous pump:

\begin{eqnarray}\label{Apavg}
    \delta_p \: \langle \big| \textsf{A}_p\big|^2 \rangle &\approx&
    \tau_f  \: I_p  \ ,
\end{eqnarray}

\noindent which is identical to Eq. (\ref{Apavg p}), only with
$\tau_f$ replacing $\tau_p$. In other words, although the pump is
continuous, we can treat it as if it was composed of serious of
pulses, each of them $\tau_f$ long. Since the atomic state, which
is the slowest component in our system, has a "memory" only
$\tau_f$ long, its response is not affected by interactions that
occurred more than $\tau_f$ seconds ago. All the other components
of our system have shorter coherence time; for example, since the
pulse bandwidth is significantly wider than
$\gamma_f=2\pi/\tau_f$, it is not affected by such 'chopping' of
continuous light to pulses $\tau_f$ long. Using Eq. (\ref{Apavg}),
and assuming the final state is included within the pump spectrum,
we get:

\begin{eqnarray}
\label{IqcTPAapp}  I^c(t-\tau_i,\tau_i-\tau_s) &\approx&
\kappa_\textsc{tpa} \: \frac{\tau_f}{\delta_p} \:
\textsc{u}(t-\tau_i) \: e^{-2\gamma_f(t-\tau_i)} \: \Delta^2\:
(n^2+n) \: P_e(\tau_i-\tau_s) \ ,
\end{eqnarray}

\noindent and the ratio between the coherent and the incoherent
signals is therefore:

\begin{eqnarray}
\label{ratioTPA} \frac{I^c(t-\tau_i,\tau_i-\tau_s)}{
I^{ic}(t-\tau_i,\tau_i-\tau_s) } &\approx&
\frac{I_p^2}{\int_{-\infty}^\infty
\texttt{I}_p(t'+\tau_s-\tau_i)\texttt{I}_p(t') \: dt'} \:
\frac{\Delta}{\delta_p} \  \frac{n^2+n}{n^2} \: P_e(\tau_i-\tau_s)
\ .
\end{eqnarray}

Similarly to the case with SFG, we see that the coherent signal is
stronger than the incoherent one (in the absence of dispersion or
delay between the signal and idler), this time roughly by the
ratio between the down-converted bandwidth and the pump bandwidth
$\Delta / \delta_p$.\\

To clarify the dependence of the incoherent signal on a relative
delay between the signal and the idler, let us consider the case
of a continuous, stationary pump, for which $\big\langle
\texttt{I}_p(t+\tau)\big\rangle=\big\langle
\texttt{I}_p(t)\big\rangle=I_p$, and so the intensity correlations
can be represented by the normalized second-order correlation
function of the pump:

\begin{eqnarray} \label{I > g2}
\frac{\int_{\tau_f} \texttt{I}_p(t'+\tau_s-\tau_i)\texttt{I}_p(t')
\: dt'}{I_p^2} \approx  \tau_f \: \frac{\big\langle
\texttt{I}_p(t+\tau_s-\tau_i)\texttt{I}_p(t)
\big\rangle}{\big\langle \texttt{I}_p(t+\tau_s-\tau_i)\big\rangle
\big\langle\texttt{I}_p(t) \big\rangle}= \tau_f \:
\texttt{g}^{(2)}_p(\tau_i-\tau_s) \ ,
\end{eqnarray}
\noindent where $\tau_f$ is the lifetime of the atomic $|f\rangle$
state, which is the physical time interval over which the
intensity correlations are actually integrated. Equation \ref{I
> g2} is valid as long as the lifetime of the final state is much
longer than the coherence time of the pump, which is indeed the
case considered here since we assumed $\delta_p\gg\gamma$.
Therefore we may write:

\begin{eqnarray}
\label{IcTPAapp} I^{ic}(t-\tau_i,\tau_i-\tau_s) &\approx&
\kappa_\textsc{tpa} \: \tau_f \ \texttt{g}^{(2)}_p(\tau_i-\tau_s)
\: \textsc{u}(t-\tau_i) \: e^{-2\gamma_f (t-\tau_i)} \: \Delta \:
n^2 \ ,
\end{eqnarray}

\noindent and the ratio between the coherent and incoherent
signals then becomes:

\begin{eqnarray}
\label{ratioTPA} \frac{I^c(t-\tau_i,\tau_i-\tau_s)}{
I^{ic}(t-\tau_i,\tau_i-\tau_s) } &\approx&
\frac{1}{\texttt{g}^{(2)}_p(\tau_i-\tau_s)} \:
\frac{\Delta}{\delta_p} \  \frac{n^2+n}{n^2} \: P_e(\tau_i-\tau_s)
\ .
\end{eqnarray}

If the final state is inhomogeneously broadened with an
inhomogeneous bandwidth $\gamma_\textsc{ih}$, we may use Eq.
(\ref{Itotal}) to obtain (assuming this time that
\mbox{$\gamma_\textsc{ih}
> \delta_p$}):

\begin{eqnarray}
\label{IqcTPAappIH} I_\textsc{ih}^c(t-\tau_i,\tau_i-\tau_s)
&\approx& \kappa_\textsc{tpa} \: \tau_f \: \textsc{u}(t-\tau_i) \:
e^{-2\gamma_f(t-\tau_i)} \: \Delta^2\: (n^2+n) \:
P_e(\tau_i-\tau_s) \nonumber \\[5mm]
I_\textsc{ih}^{ic}(t-\tau_i,\tau_i-\tau_s) &\approx&
\kappa_\textsc{tpa} \: \tau_f \: \texttt{g}^{(2)}_p(\tau_i-\tau_s)
\: \textsc{u}(t-\tau_i) \: e^{-2\gamma_f (t-\tau_i)} \:
\gamma_\textsc{ih} \: \Delta\: n^2 \ ,
\end{eqnarray}

\noindent and the ratio between the coherent and the incoherent
signals is then the same as in Eq. (\ref{ratioTPA}), only with
$\gamma_\textsc{ih}$ replacing $\delta_p$.\\

Similar results can be obtained for a quasi-continuous pump, i.e.
if we consider the ensemble average of non transform-limited
pulses, for which we may write (for small delays, $\tau_i-\tau_s
\ll \tau_p$):

\begin{eqnarray} \label{I > g2 p}
\bigg\langle \frac{\int_{\tau_f}
\texttt{I}_p(t'+\tau_s-\tau_i)\texttt{I}_p(t') \: dt'}{I_p^2}
\bigg\rangle \approx \tau_p \: \texttt{g}^{(2)}_p(\tau_i-\tau_s) \
.
\end{eqnarray}

This leads to the same results as with a continuous pump, with the
only difference being that $\tau_p$ is replacing $\tau_f$:

\begin{eqnarray}
\label{IqcTPAapp p} \big\langle I^c(t-\tau_i,\tau_i-\tau_s)
\big\rangle &\approx& \kappa_\textsc{tpa} \:
\frac{\tau_p}{\delta_p} \: \textsc{u}(t-\tau_i) \:
e^{-2\gamma_f(t-\tau_i)} \: \Delta^2\: (n^2+n) \:
P_e(\tau_i-\tau_s) \\
\label{IcTPAapp p} \big\langle I^{ic}(t-\tau_i,\tau_i-\tau_s)
\big\rangle &\approx& \kappa_\textsc{tpa} \: \tau_p \
\texttt{g}^{(2)}_p(\tau_i-\tau_s) \: \textsc{u}(t-\tau_i) \:
e^{-2\gamma_f (t-\tau_i)} \: \Delta \: n^2 \ ,
\end{eqnarray}

\noindent and the ratio between the average coherent and the
average incoherent signals therefore remains the same as in
\ref{ratioTPA}. Accordingly, in the case of inhomogeneous
broadening the same results hold, with $\tau_p$ replacing $\tau_f$
there as well.

\subsection{Coincidence events}
It is quite intriguing to compare the results obtained for SFG and
TPA with down-converted light to the expected rate $Rc$ of
coincidence events, i.e. the simultaneous arrival of signal and
idler photons. Typically, the coincidence rate is evaluated as
proportional to the second-order correlation function
$\texttt{g}^{(2)}(\tau)$. If the temporal response of the
coincidence detectors (and the corresponding electronics) is
slower than the coherence time of the photons ($\sim 1/\Delta$),
as is typically the case, then it is taken into account by
integrating over the gating time $Tg$ :
\begin{eqnarray} \label{Rc}
Rc \propto \int_{Tg}\texttt{g}^{(2)}(\tau)\: d\tau \ .
\end{eqnarray}

In order to obtain an approximated expression for the coincidence
rate, we will use the spectral functions
$g(\xi),f(\omega,\Omega_0)$ defined in Eq. (\ref{gf coincidence}).
Essentially, this means that we treat coincidence detection as if
it was an SFG process with a very large up-converted bandwidth
$\gamma_\textsc{uc}$:
\begin{eqnarray}
\label{Rc as SFG} \gamma_\textsc{uc}=2\:\Delta \ .
\end{eqnarray}

Intuitively speaking, such SFG process may be considered as
equivalent to coincidence detection since any pair of photons that
arrives at the crystal simultaneously (i.e. with a temporal
separation that is smaller than their coherence time $\sim
1/\Delta$) has an equal probability to be up-converted, regardless
of the frequency of the resulting up-converted photon. Although we
have previously assumed that $\gamma_\textsc{uc}\ll\Delta$ (Eq.
(\ref{g<<D})), this assumption was made only to allow the neglect
of the spectral variations of $n_{s,i}(\omega)$. Therefore, if we
limit our discussion to down-converted spectrum that its average
is approximately smooth, we may use the expressions obtained for
SFG to describe $\texttt{g}^{(2)}(\tau_s-\tau_i)$, simply by
replacing $\gamma_\textsc{uc}$ by $2\Delta$:
\begin{eqnarray}
\label{coincidence 0}   \texttt{g}^{(2)}(\tau_s-\tau_i) &\propto&
 \frac{\texttt{I}_p(t-\tau_i)}{I_p} \:
\Delta^2\: (n^2+n) \: P_e(\tau_i-\tau_s)  \nonumber \\  &+&
\frac{\texttt{I}_p(t-\tau_i) \texttt{I}_p(t-\tau_s)}{I_p^2} \: 2
\: \Delta^2 \: n^2 \ ,
\end{eqnarray}
\noindent where the first term represents the coherent
contribution, and the second represents the incoherent one.
However, $\texttt{g}^{(2)}(\tau_s-\tau_i)$ represents the actual
coincidence detection rate only for infinitely fast detectors; for
broadband radiation the temporal resolution of the coincidence
detectors is typically orders of magnitude longer than the
coherence time of the photons. Therefore, applying Eq. (\ref{Rc})
and taking $Tg\gg 1/\Delta, |\tau_s-\tau_i|$, we obtain:
\begin{eqnarray}
\label{Rcc Tg}   Rc^c &\propto&
 \frac{\texttt{I}_p(t-\tau_i)}{I_p} \:
\Delta\: (n^2+n) \: P_e(\tau_i-\tau_s) \nonumber \\ \label{Rcic
Tg} Rc^{ic}&\propto& \frac{\texttt{I}_p(t-\tau_i)
\texttt{I}_p(t-\tau_s)}{I_p^2} \: 2 \: Tg\:\Delta^2 \: n^2 \ .
\end{eqnarray}

As with TPA or SFG if a temporal averaging is performed for the
case of a continuous pump, or an ensemble average is performed
with quasi-continuous, long pump pulses, the ratio
$\big\langle{\texttt{I}_p(t-\tau_i)}/{I_p}\big\rangle$ in the
coherent term approaches 1, and the ratio $\big\langle
{\texttt{I}_p(t-\tau_i) \texttt{I}_p(t-\tau_s)}/{I_p^2}
\big\rangle$ in the incoherent term approaches
$\texttt{g}^2_p(\tau_i-\tau_s)$ (for small delays). Thus the ratio
between the average coherent and the incoherent contributions
becomes:
\begin{eqnarray} \label{ratio Rc}
 \frac{Rc^{c}}{Rc^{ic}} &\approx&
\frac{1}{\texttt{g}^2_p(\tau_i-\tau_s)}\frac{1}{2 Tg \: \Delta} \
\frac{(n^2+n) \: P_e(\tau_i-\tau_s) }{n^2} \ .
\end{eqnarray}

Since at any power level down-converted light is essentially
composed of simultaneously-created photon pairs, it is natural to
assume that it will exhibit high degree of bunching, in the sense
that there will always be a significantly higher rate of
simultaneous arrivals of photons from the signal and the idler
beams (at zero delay), as compared to Poissonian or even thermal
distributions. However (and counter-intuitively), as is evident
from Eqs. (\ref{Rcc Tg})-(\ref{ratio Rc}), since $Tg\gg 1/\Delta$,
at high power levels ($n\geq 1$) the coincidence rate is dominated
by the incoherent term, which exhibits similar bunching properties
as those of the pump; thus, unlike the frequency-selective
processes of SFG and TPA, the coherent contribution to the
coincidence rate is dominant only at the very low power levels of
$n\ll1$, where down-converted light can be described as a stream
of entangled photon pairs. \\

\section{Summary and conclusions} \label{Summary}

In this paper we derived expressions for two-photon interactions
induced by broadband down-converted light that was pumped by a
narrowband laser. In section \ref{motion} we solved the equations
of motion for the annihilation and creation operators of broadband
down-converted light generated by an arbitrary narrowband pump. In
section \ref{operators} we formulated operators that represent the
photon flux or the probability amplitude of weak two-photon
interactions (i.e. assuming low efficiency of the interaction, so
that the inducing fields are not depleted) induced by arbitrary
broadband light. In section \ref{evaluation} we combined the
results of the previous sections to obtain expressions for the
intensities of two-photon interactions, namely SFG, TPA and
coincidence events, induced by broadband down-converted light, and
in section \ref{results} we explored their temporal and spectral
behaviors under various conditions.\\

Our calculations show that the intensity of two-photon
interactions induced by broadband down-converted light can be
represented as the sum of two terms, one ($I^c$) that exhibits a
coherent behavior, and a second one ($I^{ic}$) that exhibits an
incoherent behavior:
\begin{eqnarray}
I^{total}=I^c+I^{ic} \ .
\end{eqnarray}

The two terms vary dramatically both in their spectral properties,
as well as in their temporal properties. We considered the case
where the signal and the idler may propagate freely along
different optical paths from the down-converting crystal,
accumulating independent temporal delays $\tau_s$, $\tau_i$ before
inducing the two-photon interaction. The coherent signal then
responds to a relative delay between the signal and the idler in
an ultrashort-pulse like behavior:
\begin{eqnarray}
I^c(t,\tau_s,\tau_i) \propto \texttt{I}_p(t-\tau_i) \:
P_e(\tau_i-\tau_s) \ ,
\end{eqnarray}

\noindent where $\texttt{I}_p(t)$ is the power of the pump (in
units of photon flux), and $P_e(\tau_i-\tau_s)$ is the temporal
response one would have got if the two-photon interaction was
induced by mixing two ultrashort, transform-limited pulses with
the same power spectra as the signal and the idler, although the
signal and the idler are each incoherent and may even be
continuous (see Fig. \ref{Fig temporal}(c)). Accordingly,
$P_e(\tau_i-\tau_s)$ is sensitive to dispersion just as a coherent
ultrashort pulse (including the dispersion of the down-conversion
process itself), and can even be shaped by conventional
pulse-shaping techniques (see Figs. \ref{Fig temporal}(d)-(f)).
Note that at low powers this corresponds to shaping of the
second-order correlation function $\texttt{g}^{(2)}$ of the
down-converted entangled photon pairs. It is also interesting to
note that $P_e(\tau_i-\tau_s)$ responds to the antisymmetric sum
of the phases applied to the signal and the idler:
$\theta_s(\omega)+\theta_i(\langle \omega_p \rangle -\omega)$,
with $\langle \omega_p \rangle$ being the center frequency of the
pump. Thus, if the same spectral filter is applied to both the
signal and the idler beams, only phase functions that are
symmetric about $\langle \omega_p \rangle /2$ affect $P_e$.
Similarly, if the signal and the idler travel through the same
medium, only odd orders of dispersion will have
an effect on $P_e$.\\

In contrast, the incoherent signal depends only on the temporal
overlap between the intensities of the down-converted signal and
idler beams, and so reacts to a delay between the signal and the
idler on the same time-scale as the long pulses or even continuous
behavior of the pump:

\begin{eqnarray}
I^{ic}(t,\tau_s,\tau_i) \propto \texttt{I}_p(t-\tau_i)
\texttt{I}_p(t-\tau_s)
\end{eqnarray}
if the pump is narrower than the final state, or
\begin{eqnarray}
I^{ic}(t,\tau_s,\tau_i) \propto \int_{-\infty}^\infty
\texttt{I}_p(t'+\tau_s-\tau_i)\texttt{I}_p(t') \: dt' \ ,
\end{eqnarray}
if the final state is narrower than the pump. In both cases, if we
consider the temporal average in the case of a continuous pump, or
the ensemble average in the case of a quasi-continuous pump (i.e.
non-transform limited pulses, for which $\tau_p\gg2\pi/\delta_p$,
with $\tau_p$ being the duration of the pulses, and $\delta_p$
their bandwidth), then the average incoherent signal is
proportional to the normalized second-order correlation function
of the pump (for $\tau_i-\tau_s \ll \tau_p$):
\begin{eqnarray}
\big \langle I^{ic}(t,\tau_s,\tau_i) \big\rangle \propto
\texttt{g}^2_p(\tau_i-\tau_s) \ .
\end{eqnarray}

Thus, as depicted in Fig. \ref{Fig temporal}, there are three
temporal timescales in our system. The longest one is the duration
of the pump pulse (which can be infinity for a continuous pump).
This timescale dictates the temporal behavior of the incoherent
signal as a function of the signal-idler delay. The next is the
coherence time of the pump which is $\sim 1/\delta_p$ (and is
equal to the duration of the pump pulse, in case it is a
transform-limited one). For signal-idler delays which are shorter
than this coherence time, the average of the incoherent signal is
higher since the intensities of the signal and the idler become
correlated, as they both reflect the intensity fluctuations of the
pump. The shortest time scale is the behavior of the coherent
signal, which is on the same timescale as the coherence time of
the broadband down-converted light: $1/\Delta$.\\

As for the spectral behavior, the coherent signal behaves as
though the interaction is actually being induced by the pump
itself, and not by the down-converted light. Thus, the coherent
signal is induced only if the pump spectrum overlaps with the
final state:
\begin{eqnarray}
I^c(\Omega_0+\xi) \propto \bigg| \int_\gamma d\xi \: g(\xi) \:
\textsf{A}_p(\Omega_0+\xi) \: \bigg|^2 \ ,
\end{eqnarray}
\noindent where $g(\xi)$ represents the the spectrum of the final
atomic level in TPA, or the phase-matching function for
up-conversion in the case of SFG, and with $\Omega_0$ being the
center frequency of the final atomic level or of the phase-matched
spectrum in case of SFG. The consequences of this spectral
behavior is that by scanning the pump wavelength we can perform
two-photon spectroscopy with the spectral resolution of the
narrowband pump, even though the interaction is induced by light
that is orders of magnitude wider than the pump, and not by the
pump itself (see Fig. \ref{Fig TPA spectrum}). In the case of SFG
this means that light is being up converted only at those
wavelengths:
\begin{eqnarray}
I^c_\textsc{sfg}(\Omega_0+\xi) \propto \big| g(\xi) \:
\textsf{A}_p(\Omega_0+\xi) \: \big|^2 \ ,
\end{eqnarray}
so that even if the phase matching conditions allow broadband
up-conversion, $I^c_\textsc{sfg}$ replicates the narrow spectrum
of the pump (see Fig. \ref{Fig SFG spectrum}).\

The incoherent signal, on the other hand, is insensitive to the
exact wavelength of the pump that generated the down-converted
light. Since the information on the original wavelength of the
pump is imprinted in the phase correlations between the
down-converted modes, it affects only the coherent signal $I^c$.
Accordingly, the incoherent signal is induced at all the possible
frequency band of the final state of the interaction (see Fig.
\ref{Fig SFG spectrum}):
\begin{eqnarray}
I^{ic}(\Omega_0+\xi)\propto \big| g(\xi)\big|^2 \ .
\end{eqnarray}


The coherent and incoherent signals also exhibit different
dependencies on $n$, the average photon-flux spectral density, and
on the bandwidth $\Delta$ of the down-converted light. While the
incoherent signal depends quadratically on $n$, the coherent
signal includes an additional, non-classical term that depends
linearly on $n$:
\begin{eqnarray}
I^c &\propto& (n^2+n)  \
\\
I^{ic} &\propto&  n^2 \ .
\end{eqnarray}

This behavior is presented in Fig. \ref{Fig LinSFG}.\\

Additionally, since the coherent signal results from coherent
summation over the entire (correlated) spectra of the signal and
the idler, it depends quadratically on $\Delta$, while the
incoherent signal depends only linearly on $\Delta$\\

Thus, excluding the case of pump pulses that are
transform-limited, the ratio between the average coherent and
incoherent signals can be represented as:

\begin{eqnarray}
 \frac{\big\langle I^c \big\rangle}{\big\langle I^{ic} \big\rangle} &\approx&
 \frac{1}
 {\texttt{g}^2_p(\tau_i-\tau_s)}
 \: \frac{\Delta}{\textrm{max}\big(\gamma,\delta_p\big)} \:
\frac{(n^2+n)}{n^2} \: P_e(\tau_i-\tau_s) \ ,
\end{eqnarray}

\noindent where $\gamma$ is the bandwidth of the final state. As
long as the delay between the signal and the idler beams is
smaller than $1/\Delta$, and in the absence of odd-order
dispersion, we can assign $P_e \approx 1$. Taking into account
that typically $1\leq \texttt{g}^2_p(0) \leq 2$, we see that the
coherent signal is dominant not only at low photon fluxes ($n\ll
1$, i.e. at the entangled-photons regime) but also at
classically-high power levels, as long as both the pump and the
final state of the interaction are narrower than the
down-converted bandwidth:
\begin{eqnarray}
 \frac{\big\langle I^c \big\rangle}{\big\langle I^{ic} \big\rangle} &\approx&
\frac{\Delta}{\textrm{max}\big(\gamma,\delta_p\big)} \:
\end{eqnarray}

In the case of coincidence detection, the relatively long gating
time $T_g$ of the electronic coincidence detection circuit makes
the incoherent contribution to the coincidence counts rate $Rc$
much larger:
\begin{eqnarray}
\frac{Rc^{c}}{Rc^{ic}} &\approx& \frac{1}{2 Tg \: \Delta} \:
\frac{(n^2+n)}{n^2} \ .
\end{eqnarray}

Thus, for the coherent contribution to dominate in electronic
coincidence-detection, one is restricted to very low photon fluxes
($n \ll 1 $). \\

It is important to note that in this paper we took into account
only two-photon interactions that result from cross-mixing of the
signal and the idler fields and not from self-mixing of the signal
with itself or the idler with itself. In both TPA and SFG, the
cross-mixing term can be isolated spectrally if the signal and the
idler are non-degenerate. In SFG, the cross-mixing term can also
be isolated spatially if the down-conversion is non-collinear.
However, in cases where the self-mixing term is indistinguishable
from the cross-mixing terms (for example in the case of TPA with
degenerate signal and idler fields, or if degenerate and collinear
down conversion is considered) this has the effect of increasing
the incoherent signal by a factor of two:
\begin{eqnarray} I^{ic}\Rightarrow 2I^{ic} \end{eqnarray}

Naturally, the coherent signal is generated only by cross-mixing
of the signal and the idler fields, and therefore is not affected
by such self-mixing terms.\\

Finally we note again that none of the effects described in this
paper is directly related to squeezing. Even the non-classical
linear intensity dependence is in fact independent of squeezing;
since the coherent and incoherent signals are attenuated equally
(quadratically) by such losses, this effect can be observed even
in the presence of losses that would wipe out the squeezing
properties completely. Moreover, while the squeezing degree grows
with $n$ and is very small for $n \leq 1$, the linear term becomes
less and less dominant as $n$ grows, and is completely negligible
at $n \gg 1$. Furthermore, excluding the linear intensity
dependence of the coherent signal, all the other effects
considered in this paper are completely described within the
classical framework. Indeed, such effects can be created by
appropriately shaping classical pulses, so that they obtain
similar anti-symmetric spectral phase correlations
\cite{Salehi@Herttage_JLT_1990,{Meshulach@Silberberg_PRA_1999}}.
However, the precision of these correlations in broadband
down-converted light can be many orders of magnitude higher than
achievable by pulse-shaping techniques
\cite{Peer@Friesem_JLT_2004,{Peer@Friesem_PRA_2006}}. The unique
properties of two-photon interactions induced by broadband
down-converted light are therefore both interesting and
applicable.

\begin{acknowledgments}
I wish to thank Avi Pe'er and Yaron Silberberg for many fruitful
discussions and insights.
\end{acknowledgments}

\end{document}